\pdfoutput=1
\RequirePackage[T1]{fontenc}
\documentclass[12pt]{article}

\usepackage[T1]{fontenc}
\usepackage[utf8]{inputenc}
\usepackage{microtype}

\usepackage{float}
\usepackage{multicol}
\usepackage{bm}
\usepackage{booktabs}
\usepackage{siunitx}
\usepackage{multirow}
\usepackage{amsmath,amssymb,amsfonts,amsthm,bbm}

\usepackage{mathtools}
\usepackage{graphicx}
\usepackage{textcomp}
\usepackage[table]{xcolor}
\usepackage{nicefrac}
\usepackage{braket}
\usepackage{adjustbox}
\usepackage{rotating}
\usepackage[percent]{overpic}
\usepackage{longtable}
\usepackage{algorithm}
\usepackage{algorithmicx}
\usepackage{algpseudocode}
\usepackage{thmtools,thm-restate}
\usepackage{enumitem}
\usepackage{authblk}
\usepackage{subcaption}
\usepackage[pagebackref=true,breaklinks=true,colorlinks,hyperfootnotes=false]{hyperref}
\hypersetup{
  colorlinks,
  citecolor=citeblue,
  linkcolor=firebrick,
  urlcolor=firebrick
  }
\usepackage[nameinlink,capitalize,noabbrev]{cleveref}

\usepackage{fullpage}
\usepackage[sort]{natbib}
\usepackage{orcidlink}
\usepackage{fancyhdr}
\newcommand{\runningtitle}{Rethinking Quantum Continual Learning with Quantum
Fisher Information}
\def\BibTeX{{\rm B\kern-.05em{\sc i\kern-.025em b}\kern-.08em
    T\kern-.1667em\lower.7ex\hbox{E}\kern-.125emX}}

\DeclareFontFamily{OT1}{pzc}{}
\DeclareFontShape{OT1}{pzc}{m}{it}{<-> s * [1.10] pzcmi7t}{}
\DeclareMathAlphabet{\pzccal}{OT1}{pzc}{m}{it}

\definecolor{lightyellow}{rgb}{1.0, 0.95, 0.7}
\definecolor{Blue}{rgb}{0, 0, 0.8}
\definecolor{blue}{rgb}{0,0,1}
\definecolor{mydarkblue}{rgb}{0,0.08,0.45}
\definecolor{mydarkblue2}{rgb}{0.133, 0.133, 0.698}
\definecolor{echodrk}{HTML}{0099cc}
\definecolor{mymauve}{rgb}{0.58,0,0.82}
\definecolor{darkgreen}{rgb}{0,0.40,0}
\definecolor{firebrick}{rgb}{0.698,0.133,0.133}
\definecolor{midnightblue}{rgb}{0.1,0.1,0.44}
\definecolor{citeblue}{RGB}{0, 113, 188}
\definecolor{oxfordblue}{rgb}{0.0,0.13,0.28}
\definecolor{prussianblue}{rgb}{0.0,0.19,0.33}
\definecolor{coolteal}{rgb}{0, 0.45, 0.45}
\definecolor{olive}{rgb}{0.1, 0.3, 0}
\definecolor{mypurple}{rgb}{0.5,0,0.5}
\definecolor{almond}{rgb}{0.94, 0.87, 0.8}

\definecolor{blue_ampEncoding}{HTML}{DAE8FC}
\definecolor{green_encoder}{HTML}{D5E8D4}
\definecolor{purple_decoder}{HTML}{E1D5E7}
\definecolor{yellow_measure}{HTML}{FFF2CC}
\definecolor{gray_block}{HTML}{F5F5F5}
\definecolor{pink_dru}{HTML}{FAD9D5}
\definecolor{orange_v}{HTML}{FAD7AC}
\definecolor{Lightblue}{HTML}{E7F4FC}

\DeclareDocumentCommand \norm { o m }{{\lVert #2 \rVert_#1}}

\declaretheoremstyle[%
  spaceabove=10pt,%
  spacebelow=2pt,%
  headfont=\normalfont\itshape,%
  postheadspace=0em,%
  qed=%
]{prfstyle}

\title{\LARGE \rmfamily \bfseries
Rethinking Quantum Continual Learning with Quantum Fisher Information}

\author[1,2,3,\orcidlink{0009-0004-7221-3854}]{Yu-Chao Hsu}
\author[4,\orcidlink{0009-0009-1911-9566}]{Yu-Cheng Lin}
\author[3,\orcidlink{0000-0002-1993-1863}]{\authorcr Tai-Yue Li}
\author[3,\orcidlink{0000-0001-8139-6809}]{Nan-Yow Chen}
\author[4,\orcidlink{0000-0000-0000-0000}]{En-Jui Kuo\textsuperscript{$\dagger$,}}

\affil[1]{School of Electrical Engineering, Korea Advanced Institute of Science and Technology (KAIST), 291 Daehak-ro, Yuseong-gu, Daejeon 34141, Republic of Korea}
\affil[2]{Cross College Elite Program, National Cheng Kung University, Tainan 701401, Taiwan}
\affil[3]{National Center for High-Performance Computing , National Institutes of Applied Research (NIAR), Hsinchu 300092, Taiwan}
\affil[4]{Department of Electrophysics, National Yang Ming Chiao Tung University, Hsinchu 300092, Taiwan.}

\affil[$\dagger$]{Correspondence to: 
\href{mailto:kuoenjui@nycu.edu.tw}{\texttt{kuoenjui@nycu.edu.tw}}.
}
\date{\today}

\begin{document}
\maketitle

\begingroup
\renewcommand\thefootnote{}
% \footnotetext{The views expressed in this article are those of the authors and do not represent the views of Wells Fargo. This article is for informational purposes only. Nothing contained in this article should be construed as investment advice. Wells Fargo makes no express or implied warranties and expressly disclaims all legal, tax, and accounting implications related to this article.}
\endgroup

\clearpage

\begin{abstract}
Quantum continual learning seeks to enable quantum learning models to acquire sequential tasks while retaining previously learned knowledge. 
However, variational quantum classifiers (VQCs) remain vulnerable to catastrophic forgetting when trained under nonstationary task distributions. 
In this work, we propose quantum elastic weight consolidation (QEWC), a quantum Fisher information (QFI)-informed regularization framework for mitigating forgetting in quantum continual learning. 
In contrast to conventional elastic weight consolidation based on classical Fisher information (CFI), which defines parameter importance through measurement-dependent output statistics, QEWC uses the QFI to quantify the intrinsic sensitivity of the parameterized quantum state. 
This formulation provides an information-geometric perspective in which parameter importance is determined by the local response of the quantum state manifold to parameter variations. 
We evaluate QEWC using VQCs trained on sequential binary classification tasks, including classical image-classification tasks and a quantum phase-classification task. 
Our simulations show that unregularized sequential training leads to severe catastrophic forgetting, whereas both CFI-based EWC and QFI-based QEWC substantially improve the retention of previously learned tasks. 
Further mechanistic analyses reveal that the two approaches induce distinct regularization geometries. 
The CFI-based penalty acts more selectively on measurement-sensitive parameter directions, whereas the QFI-based penalty imposes a denser state-geometric constraint over the parameter space.
This distinction leads to different stability--plasticity behaviors during sequential training. 
Under depolarizing noise, the CFI values are strongly suppressed by degraded measurement statistics, while the QFI retains a more stable sensitivity structure associated with the noisy parameterized quantum state.
These results establish QEWC as a physically motivated framework for studying and mitigating forgetting in quantum continual learning through the geometry of the underlying quantum state.

\vspace{1em}
\end{abstract}

\section{Introduction}
Humans and animals can continually acquire new experiences and skills while preserving previously acquired knowledge. Inspired by this biological capability, continual learning, also referred to as incremental or lifelong learning, seeks to endow artificial intelligence with the ability to continually adapt to evolving environments without losing prior knowledge~\cite{10.5555/3086758,10.1109/TPAMI.2024.3367329,PARISI201954}.
More specifically, continual learning addresses the challenge of learning from non-stationary data distributions that evolve over time, without assuming that the number or order of tasks is known in advance~\cite{PARISI2017137}. However, achieving this goal remains highly challenging in artificial learning systems. A central obstacle is catastrophic forgetting, where learning a new task often leads to a substantial degradation in performance on previously learned tasks~\cite{MCCLOSKEY1989109,goodfellow2015}. 
% This difficulty becomes particularly pronounced when the tasks are dissimilar or when the model is trained over long task sequences.
This phenomenon reflects a fundamental trade-off between learning plasticity and memory stability~\cite{Kirkpatrick2017Overcoming,kemker2018measuring,Dohare2024,chen2025Intrinsic}: the model must remain sufficiently plastic to acquire new knowledge while being stable enough to preserve previously learned information. In recent years, several attempts have been made to tackle this problem, and continual learning has consequently developed into a rapidly growing research field with promising applications across a wide range of domains, including robotics~\cite{pmlr-v232-powers23a,10802683}, financial markets~\cite{philps2019Continual} , and human activity recognition~\cite{JHA20211}.

On the other hand, over the past few decades, quantum computing~\cite{Arute2019,Daley2022,RevModPhys.93.025001} has emerged as a promising paradigm for information processing, owing to its potential advantages in solving certain computational tasks beyond the reach of classical approaches, therefore motivating a broad range of quantum machine learning (QML) algorithms, including quantum neural networks
~\cite{Benedetti_2019,Abbas2021,Beer2020,Liu2025,11249855}, quantum kernel methods~\cite{Havlicek2019,PhysRevLett.122.040504,Chen_2025Validating,huang2026}, quantum convolutional neural networks~\cite{Cong2019Quantum,11000235}, and quantum eigensolver~\cite{Xia_2021,Crognaletti_2025,Moll_2018,lin2026generative}
, some of which are particularly compatible with noisy intermediate-scale quantum (NISQ)~\cite{Preskill2018} devices and have demonstrated advantages over classical methods in certain tasks. However, despite these advances, most existing quantum learning models
are still designed for specific predefined tasks and static data distributions. 
In contrast, the problem of enabling quantum models to learn continually from sequentially arriving tasks has only recently started to be explored~\cite{chen2025Intrinsic,Jiang_2022Quantum,Zhang2026Experimental,10448379,zhu2026qclids,ZHENG2026134086}. Recent work has further demonstrated quantum continual learning on real quantum devices~\cite{Zhang2026Experimental}.

Among these emerging efforts, one representative theoretical work extends the standard elastic weight consolidation (EWC) framework~\cite{Kirkpatrick2017Overcoming} to the setting of quantum continual learning~\cite{Jiang_2022Quantum}.
However, how to quantify parameter importance in quantum continual learning remains an open question. Although existing EWC-based approaches provide an important starting point, they largely inherit the classical notion of parameter importance from conventional machine learning, most notably the classical Fisher information (CFI). This choice is not entirely natural for quantum learning models, because the CFI depends explicitly on the measurement being performed and therefore only characterizes parameter sensitivity with respect to a specific measurement scheme.
For parametrized quantum learning models, the underlying object being optimized is a quantum state generated by a parameterized quantum circuit. Accordingly, the quantum Fisher information (QFI) provides a more intrinsic, measurement-independent measure of parameter sensitivity than the CFI, as it quantifies the local distinguishability of neighboring quantum states induced by infinitesimal parameter variations~\cite{Braunstein1994,Liu_2020,Meyer2021fisherinformationin,Cerezo_2021}. 
More precisely, for any fixed positive-operator-valued measure (POVM), the corresponding CFI matrix is bounded by the QFI matrix in the L\"owner order. The QFI therefore provides a measurement-independent upper bound on the local statistical distinguishability encoded in the parametrized quantum state. Owing to this measurement-independent nature, the QFI has played a central role in quantum metrology~\cite{PhysRevA.91.042104,Lu2015,PhysRevResearch.4.L012014}, quantum steering~\cite{PhysRevA.105.022421,PhysRevResearch.5.013103}, and quantum estimation~\cite{PRXQuantum.2.020303,PRXQuantum.2.010343}. Motivated by these observations, we revisit continual learning in quantum models from the perspective of QFI and develop the QFI-informed regularization framework for mitigating catastrophic forgetting.

In this work, we propose quantum elastic weight consolidation (QEWC), a QFI-informed regularization framework for quantum continual learning. Rather than treating parameter importance as a purely classical and measurement-dependent quantity, the QEWC framework offers an information-geometric perspective on catastrophic forgetting in quantum learning models. Specifically, by using the QFI as the regularization metric, the QEWC framework quantifies how variations in trainable parameters affect the underlying quantum state itself, allowing previously acquired knowledge to be preserved during sequential training in a measurement-independent manner.
To demonstrate the effectiveness of this perspective, we construct a series of variational quantum classification tasks. Our results show that quantum classifiers can suffer from severe catastrophic forgetting under sequential training, while both CFI- and QFI-based regularization substantially mitigate this effect. Importantly, the QFI-based formulation provides a more intrinsic characterization of parameter sensitivity in quantum learning models.

\section{Quantum Continual Learning}
Quantum neural networks (QNNs), typically implemented by parameterized quantum circuits (PQCs), have emerged as a promising framework for quantum machine learning (QML). 
Most existing QML models, however, are designed for isolated tasks with fixed data distributions. 
In realistic learning scenarios, a quantum model may instead encounter a sequence of tasks with changing data distributions, requiring it to acquire new knowledge while retaining what has been learned previously. 
This gives rise to the setting of quantum continual learning.

Similar to their classical counterparts, QNNs are susceptible to \emph{catastrophic forgetting}. 
When a QNN is trained sequentially on multiple tasks, optimizing the model for a newly encountered task can substantially degrade its performance on earlier ones. 
Formally, let $\{T_1,T_2,\ldots,T_K\}$ denote a sequence of learning tasks, each associated with a loss function $\mathcal{L}_k(\boldsymbol{\theta})$. 
After training on task $T_k$, the model parameters are optimized toward a task-specific solution $\boldsymbol{\theta}_k^{*}$. 
Subsequent training on $T_{k+1}$ may drive the parameter vector away from regions that are important for the previously learned tasks and lead to performance degradation on earlier tasks.

\subsection{Elastic Weight Consolidation}

To mitigate catastrophic forgetting in continual learning, regularization-based approaches such as Elastic Weight Consolidation (EWC) framework has been widely adopted in classical continual learning. 
Following the general framework for quantum continual learning proposed in Ref.~\cite{Jiang_2022Quantum}, we implement the EWC framework and adapt it to QNNs, where the model parameters are optimized sequentially across tasks. 
This adaptation enables the preservation of knowledge acquired from earlier tasks while facilitating effective learning of new tasks in a quantum setting.

We begin by specifying the task-dependent loss function used in our experiments. 
Since we focus on binary classification, when training the $k$-th task $T_k$ with dataset $\mathcal{D}_k$, we adopt the binary cross-entropy loss between the ground-truth labels $y\in\{0,1\}$ and the predicted class probabilities produced by the VQC. 
The empirical loss for task $T_k$ is given by
\begin{equation}
\mathcal{L}_k(\boldsymbol{\theta})
=
-\frac{1}{|\mathcal{D}_k|}
\sum_{(x,y)\in \mathcal{D}_k}
\log p_{\boldsymbol{\theta}}(y|x),
\label{eq:bce_loss}
\end{equation}
where $p_{\boldsymbol{\theta}}(y|x)$ denotes the predicted probability assigned by the VQC to the target label $y$ given input $x$.

To preserve knowledge acquired from previously learned tasks, the EWC framework augments the loss function for the current task with a quadratic consolidation penalty. 
Suppose that the model is trained sequentially on a sequence of tasks $\{T_1,T_2,\ldots,T_K\}$. 
When learning the $k$-th task, the EWC objective is defined as
\begin{equation}
\mathcal{L}_{\mathrm{EWC}}^{(k)}(\boldsymbol{\theta})
=
\mathcal{L}_k(\boldsymbol{\theta})
+
\frac{\lambda}{2}
\sum_{j=1}^{k-1}
(\boldsymbol{\theta}-\boldsymbol{\theta}_{j}^{*})^\top
F^{(j)}_{\mathrm C}(\boldsymbol{\theta}_{j}^{*})
(\boldsymbol{\theta}-\boldsymbol{\theta}_{j}^{*}),
\label{eq:ewc_general_1}
\end{equation}
where $\mathcal{L}_k(\boldsymbol{\theta})$ is the task loss for $T_k$, $\boldsymbol{\theta}_{j}^{*}$ denotes the parameter vector obtained after training on the previous task $T_j$, and $F^{(j)}_{\mathrm C}(\boldsymbol{\theta}_{j}^{*})$ is the task-dependent CFI matrix evaluated at $\boldsymbol{\theta}_{j}^{*}$ using the data distribution of task $T_j$. 
The CFI assigns larger weights to parameters that are more important for the previously learned task, so deviations along these directions are penalized more strongly during subsequent training.

The hyperparameter $\lambda$ controls the strength of this consolidation penalty and therefore mediates the stability--plasticity trade-off. 
A larger value of $\lambda$ enforces stronger preservation of past-task parameters, improving memory stability but potentially reducing the model's flexibility in adapting to the new task. 
Conversely, a smaller value allows greater plasticity for the current task, but may lead to more severe forgetting. 
Consequently, the EWC framework mitigates catastrophic forgetting by selectively restricting updates along parameter directions that are important for previously acquired knowledge.

\subsection{Variational Quantum Classifier}
\label{sec:VQC}
We employ a variational quantum classifier (VQC) as the underlying quantum model for sequential binary classification tasks. 
The classifier consists of three main components: an amplitude-encoding circuit, a trainable variational circuit, and a measurement-based readout. 
In this architecture, a classical input vector is first embedded into the amplitudes of an $n$-qubit quantum state, subsequently processed by a parametrized quantum circuit, and finally converted into class probabilities through measurements of predefined observables, such as Pauli-$Z$ operators. 
This provides a natural platform for studying catastrophic forgetting in quantum continual learning, since sequential training on newly arriving tasks may move the circuit parameters away from configurations that are important for previously learned tasks.

Let $\mathcal{X}\subseteq\mathbb{R}^{2^n}$ denote the input space, and let $\mathbf{x}=(x_0,x_1,\ldots,x_{2^n-1})^{\mathsf T}\in\mathcal{X}$ be a normalized input feature vector satisfying $\|\mathbf{x}\|_2=1$. 
We denote the data-encoding circuit by $U_{\mathrm{enc}}(\mathbf{x})$ and the trainable variational circuit by $U_{\mathrm{VQC}}(\boldsymbol{\theta})$, where $\boldsymbol{\theta}$ collects all trainable parameters of the quantum classifier.
In our implementation, the input vector is encoded by amplitude encoding. 
More specifically, the encoding circuit prepares the $n$-qubit state
\begin{equation}
U_{\mathrm{enc}}(\mathbf{x})|0\rangle^{\otimes n}
=
|\mathbf{x}\rangle
=
\sum_{b=0}^{2^n-1}
x_b |b\rangle ,
\label{eq:uenc}
\end{equation}
where $\{|b\rangle\}_{b=0}^{2^n-1}$ denotes the computational basis. 
Therefore, each component of the normalized input vector is represented as the amplitude of a computational basis state in the $2^n$-dimensional Hilbert space.

After data encoding, the quantum state is processed by a hardware-efficient variational circuit. 
The circuit consists of $L$ repeated variational layers, each composed of trainable single-qubit rotations followed by a nearest-neighbor CNOT entangling layer. 
Explicitly, the variational unitary is written as
\begin{equation}
U_{\mathrm{VQC}}(\boldsymbol{\theta})
=
\prod_{\ell=1}^{L}
\left[
\left(
\prod_{q=1}^{n-1}
\mathrm{CNOT}_{q,q+1}
\right)
\left(
\bigotimes_{q=1}^{n}
R_z(\theta_{\ell,q})
R_y(\theta_{\ell,q})
\right)
\right],
\label{eq:uvqc}
\end{equation}
where $\theta_{\ell,q}$ denote the trainable rotation angles applied to the $q$th qubit in the $\ell$th variational layer. 
The gate $\mathrm{CNOT}_{q,q+1}$ represents a CNOT operation with qubit $q$ as the control and qubit $q+1$ as the target. 
In this construction, the single-qubit rotations provide the trainable degrees of freedom, while the nearest-neighbor CNOT chain introduces entangling correlations across the qubit register.

The output state of the classifier is therefore given by
\begin{equation}
|\psi(\mathbf{x},\boldsymbol{\theta})\rangle
=
U_{\mathrm{VQC}}(\boldsymbol{\theta})
U_{\mathrm{enc}}(\mathbf{x})
|0\rangle^{\otimes n}.
\label{eq:vqc_state}
\end{equation}
For binary classification, we use two readout qubits to produce class-dependent scores. 
The score associated with class $c\in\{0,1\}$ is defined as the Pauli-$Z$ expectation value
\begin{equation}
z_c(\mathbf{x},\boldsymbol{\theta})
=
\langle \psi(\mathbf{x},\boldsymbol{\theta})|
\hat{Z}_c
|\psi(\mathbf{x},\boldsymbol{\theta})\rangle,
\label{eq:two_qubit_readout}
\end{equation}
where $\hat{Z}_c$ denotes the Pauli-$Z$ operator acting on the readout qubit associated with class $c$. 
These two scores are then converted into class probabilities through the softmax function,
\begin{equation}
p_{\boldsymbol{\theta}}(y=c|\mathbf{x})
=
\frac{
\exp\!\left[z_c(\mathbf{x},\boldsymbol{\theta})\right]
}{
\sum_{c'=0}^{1}
\exp\!\left[z_{c'}(\mathbf{x},\boldsymbol{\theta})\right]
}.
\label{eq:softmax_readout}
\end{equation}
The binary cross-entropy loss in Eq.~\eqref{eq:bce_loss} is then computed using $p_{\boldsymbol{\theta}}(y|\mathbf{x})$. 
The predicted label is assigned according to
\begin{equation}
\hat{y}
=
\arg\max_{c\in\{0,1\}}
p_{\boldsymbol{\theta}}(y=c|\mathbf{x}).
\label{eq:prediction_rule}
\end{equation}
Consequently, the VQC defines a trainable map from an amplitude-encoded input state to a measurement-based class prediction. 
In the continual-learning setting, the same set of circuit parameters is updated sequentially across tasks, making this architecture suitable for investigating catastrophic forgetting and forgetting mitigation.

\begin{figure*}[t]
\centering
\includegraphics[width=0.82\textwidth]{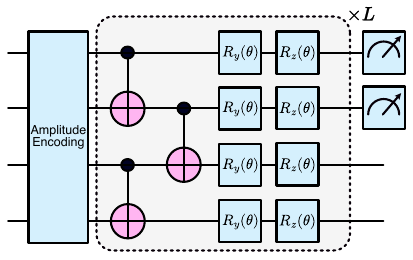}
\caption{\textbf{A $L$ layer of the variational ansatz of our quantum classifier.} All single qubit gates in this ansatz are rotation gates ($R_y(\boldsymbol{\theta})=\exp\!\left(-i\frac{\boldsymbol{\theta}}{2}\sigma_y\right)$ and $
R_z(\boldsymbol{\theta})
=
\exp\!\left(-i\frac{\boldsymbol{\theta}}{2}\sigma_z\right)
$). We measure the first and second qubit and treat its output as the classification result of this quantum classifier. Our quantum classifier used in numerical simulations consists of 30 repeated layers}
\label{fig: VQC}
\end{figure*}

\section{Quantum Fisher Information in the EWC framework}\label{sec:Theory}
\subsection{Quantum Fisher Information}

Consider a parametrized quantum state $\rho({\boldsymbol\theta})$, where $\boldsymbol{\theta}=(\theta_0,\theta_1,\ldots)^{\mathsf T}$ denotes the vector of trainable parameters. In practical quantum experiments, these parameters cannot be accessed directly. Instead, information about them is extracted from measurement statistics obtained from the quantum state. For a given POVM $\{M_a\}_a$, the probability of obtaining outcome $a$ is
$
p(a|\boldsymbol{\theta})
=
\mathrm{Tr}\!\left[\rho(\boldsymbol{\theta})M_a\right].
$
The corresponding CFI matrix is then given by
\begin{equation}
F^{\mathrm C}_{ij}
=
\sum_a
\frac{1}{p(a|\boldsymbol{\theta})}
\partial_i p(a|\boldsymbol{\theta})
\partial_j p(a|\boldsymbol{\theta}) .
\label{eq: FCij}
\end{equation}
The CFI matrix quantifies how sensitively the observed measurement statistics respond to infinitesimal changes in the parameters. However, this sensitivity is tied to the chosen measurement scheme. As a result, the CFI matrix reflects not only the local structure of the parametrized quantum state, but also the particular POVM used to extract classical information from it.

To remove this measurement dependence, one is naturally led to the QFI. In the information-geometric sense, the QFI characterizes the local distinguishability of neighboring quantum states induced by infinitesimal parameter variations. More specifically, for a fixed input $\mathbf{x}$, the VQC prepares the state
$\rho_{\mathbf{x}}(\boldsymbol{\theta})$. For any fixed POVM
$\{M_a\}_a$ satisfying $M_a\succeq 0$ and
$\sum_a M_a=\mathbb{I}$, the corresponding input-dependent CFI matrix
is bounded by the QFI matrix of this state~\cite{Liu_2020,Meyer2021fisherinformationin},
\begin{equation}
F^{\mathrm C}(\boldsymbol{\theta};\mathbf{x},\{M_a\}_a)
\preceq
F^Q\!\left[\rho_{\mathbf{x}}(\boldsymbol{\theta})\right].
\label{eq:cfi-qfi-bound}
\end{equation}

where $\preceq$ denotes the L\"owner partial order. The task-level metric used in QEWC is obtained by averaging these single-input QFI matrices over the empirical data distribution of the previously learned task. Therefore, unlike the CFI matrix associated with a fixed readout, the QFI provides a measurement-independent upper bound on the local statistical distinguishability encoded in the quantum state. We note, however, that in multiparameter settings, the simultaneous attainability of this bound generally requires additional compatibility conditions.

A standard way to define the QFI is through the symmetric logarithmic derivative (SLD) operators~\cite{HELSTROM1967101}. For each parameter $\theta_i$, the corresponding SLD operator $L_i$ is defined implicitly by
\begin{equation}
\partial_i \rho(\boldsymbol{\theta})
=
\frac{1}{2}
\left[
\rho(\boldsymbol{\theta})L_i
+
L_i\rho(\boldsymbol{\theta})
\right],
\label{eq:sld_def}
\end{equation}
where $\partial_i\equiv\partial/\partial\theta_i$. The SLD is Hermitian, and when $\rho(\boldsymbol{\theta})$ is full rank, the above equation admits a unique solution for $L_i$.

In terms of the SLD operators, the QFI matrix is given by
\begin{equation}
F^{\mathrm Q}_{ij}
=
\mathrm{Tr}\!\left[
\rho(\boldsymbol{\theta})
\frac{L_iL_j+L_jL_i}{2}
\right].
\label{eq:qfi_sld}
\end{equation}
This representation makes explicit that the QFI is a symmetric positive semidefinite matrix on the parameter space. It therefore endows the manifold of parametrized quantum states with a natural information-geometric structure.

In this work, we adopt the diagonal approximation of the task-averaged QFI matrix~\cite{Stokes2020quantumnatural,quantumAnna2025,GomezLurbe2025}. 
This choice is motivated by both practical and conceptual considerations. 
On the practical side, estimating the full QFI matrix requires access to all pairwise parameter correlations, which becomes costly for quantum learning models with many trainable parameters, as discussed in \ref{app:full}
By retaining only the diagonal entries, the QEWC regularizer assigns parameter-wise importance weights while avoiding the computational and storage overhead associated with the full QFI matrix. By contrast, the diagonal elements can be estimated more efficiently and are directly compatible with the classical EWC formulation, where each parameter is assigned an individual importance weight. On the conceptual side, the diagonal element $F^{\mathrm Q}_{ii}$ quantifies the sensitivity of the quantum state to an infinitesimal variation of the single parameter $\theta_i$, while keeping the other parameters fixed. It therefore provides a natural parameter-wise importance measure for QFI-based elastic weight consolidation.

Accordingly, each element of the full QFI matrix can be expressed in terms of the SLD operators as
\begin{equation}
F^{\mathrm Q}_{ij}
=
\mathrm{Tr}\!\left[
\rho(\boldsymbol{\theta})
\frac{L_iL_j+L_jL_i}{2}
\right],
\label{eq:qfi_sld_full}
\end{equation}
where $L_i$ and $L_j$ are the SLD operators associated with the parameters $\theta_i$ and $\theta_j$, respectively. The diagonal element $F^{\mathrm Q}_{ii}$ quantifies the sensitivity of the quantum state along a single parameter direction, while the off-diagonal elements $F^{\mathrm Q}_{ij}$ encode correlations between different parameter directions.

To obtain an explicit expression for mixed states, we write the spectral decomposition of the density operator as
\begin{equation}
\rho(\boldsymbol{\theta})
=
\sum_a
\lambda_a(\boldsymbol{\theta})
|a(\boldsymbol{\theta})\rangle
\langle a(\boldsymbol{\theta})| ,
\end{equation}
where $\lambda_a(\boldsymbol{\theta})$ denotes the $a$-th eigenvalue and
$|a(\boldsymbol{\theta})\rangle$ denotes the corresponding eigenvector.
The QFI matrix can then be expressed as
\begin{equation}
\left[F_Q(\boldsymbol{\theta})\right]_{ij}
=
2
\sum_{a,b:\lambda_a+\lambda_b>0}
\frac{
\operatorname{Re}\!\left[
\langle a|\partial_i\rho|b\rangle
\langle b|\partial_j\rho|a\rangle
\right]
}{
\lambda_a+\lambda_b
}.
\label{eq:QFI_spectral}
\end{equation}
Here, $\partial_i\rho=\partial\rho/\partial\theta_i$, and the summation is restricted to pairs $(a,b)$ satisfying $\lambda_a+\lambda_b>0$. Eq.~\eqref{eq:QFI_spectral} gives the mixed-state spectral representation of the full QFI matrix, which we use in the noisy simulations.

In the VQC considered here, the density operator also depends on the input data through the encoding circuit. We therefore apply Eq.~\eqref{eq:QFI_spectral} to the input-dependent state $\rho_{\mathbf{x}}(\boldsymbol{\theta})$ for each input $\mathbf{x}$. 
For a previously learned task $T_j$, the corresponding task-level QFI metric is defined by averaging these single-input QFI matrices over the empirical data distribution:
\begin{equation}
\bar{F}_{Q}^{(j)}(\boldsymbol{\theta}_j^\ast)
=
\frac{1}{|\mathcal{D}_j|}
\sum_{(\mathbf{x},y)\in\mathcal{D}_j}
F_Q\!\left[
\rho_{\mathbf{x}}(\boldsymbol{\theta}_j^\ast)
\right].
\label{eq:task_averaged_qfi}
\end{equation}
This task-averaged matrix quantifies the average sensitivity of the generated quantum states around the previous-task optimum $\boldsymbol{\theta}_j^\ast$.
In the diagonal implementation of QEWC, the parameter-wise importance weight is given by the diagonal element
$\left[\bar{F}_{Q}^{(j)}(\boldsymbol{\theta}_j^\ast)\right]_{ii}$.
Equivalently, this weight is obtained by computing the diagonal QFI element for each input state and then averaging it over the empirical data distribution of task $T_j$.

For pure states, Eq.~\eqref{eq:qfi_sld_full} admits a particularly simple geometric form. In the noiseless VQC setting, for a fixed input $\mathbf{x}$, the circuit prepares the rank-one state
$
\rho_{\mathbf{x}}(\boldsymbol{\theta})
=
|\psi(\mathbf{x},\boldsymbol{\theta})\rangle
\langle\psi(\mathbf{x},\boldsymbol{\theta})| .
$
The QFI matrix associated with this input-dependent pure state is then given by
\begin{equation}
\left[
F_Q\!\left[
\rho_{\mathbf{x}}(\boldsymbol{\theta})
\right]
\right]_{ij}
=
4\,\mathrm{Re}
\left[
\langle \partial_i \psi(\mathbf{x},\boldsymbol{\theta})
|
\partial_j \psi(\mathbf{x},\boldsymbol{\theta})\rangle
-
\langle \partial_i \psi(\mathbf{x},\boldsymbol{\theta})
|
\psi(\mathbf{x},\boldsymbol{\theta})\rangle
\langle
\psi(\mathbf{x},\boldsymbol{\theta})
|
\partial_j \psi(\mathbf{x},\boldsymbol{\theta})\rangle
\right],
\label{eq:qfi_pure_full}
\end{equation}
where
$\partial_i|\psi(\mathbf{x},\boldsymbol{\theta})\rangle
=
\partial|\psi(\mathbf{x},\boldsymbol{\theta})\rangle/\partial\theta_i$.
Eq.~\eqref{eq:qfi_pure_full} is equal to four times the Fubini--Study metric on the projective Hilbert space. It therefore quantifies the local distinguishability of neighboring pure states generated from the same input $\mathbf{x}$ under infinitesimal parameter variations.

Under the diagonal approximation used in our QEWC implementation, the corresponding single-input parameter sensitivity is obtained by setting $i=j$:
\begin{equation}
\left[
F_Q\!\left[
\rho_{\mathbf{x}}(\boldsymbol{\theta})
\right]
\right]_{ii}
=
4\left(
\langle \partial_i \psi(\mathbf{x},\boldsymbol{\theta})
|
\partial_i \psi(\mathbf{x},\boldsymbol{\theta})\rangle
-
\left|
\langle
\psi(\mathbf{x},\boldsymbol{\theta})
|
\partial_i \psi(\mathbf{x},\boldsymbol{\theta})\rangle
\right|^2
\right).
\label{eq:qfi_pure_diag}
\end{equation}
The task-level importance weight used by QEWC is then obtained by averaging this single-input quantity over the empirical data distribution of the previously learned task, following Eq.~\eqref{eq:task_averaged_qfi}.

\subsection{Continual Learning with Quantum Fisher Information}\label{Sec:QEWC}

\begin{figure}[t]
\centering
\includegraphics[width=0.82\textwidth]{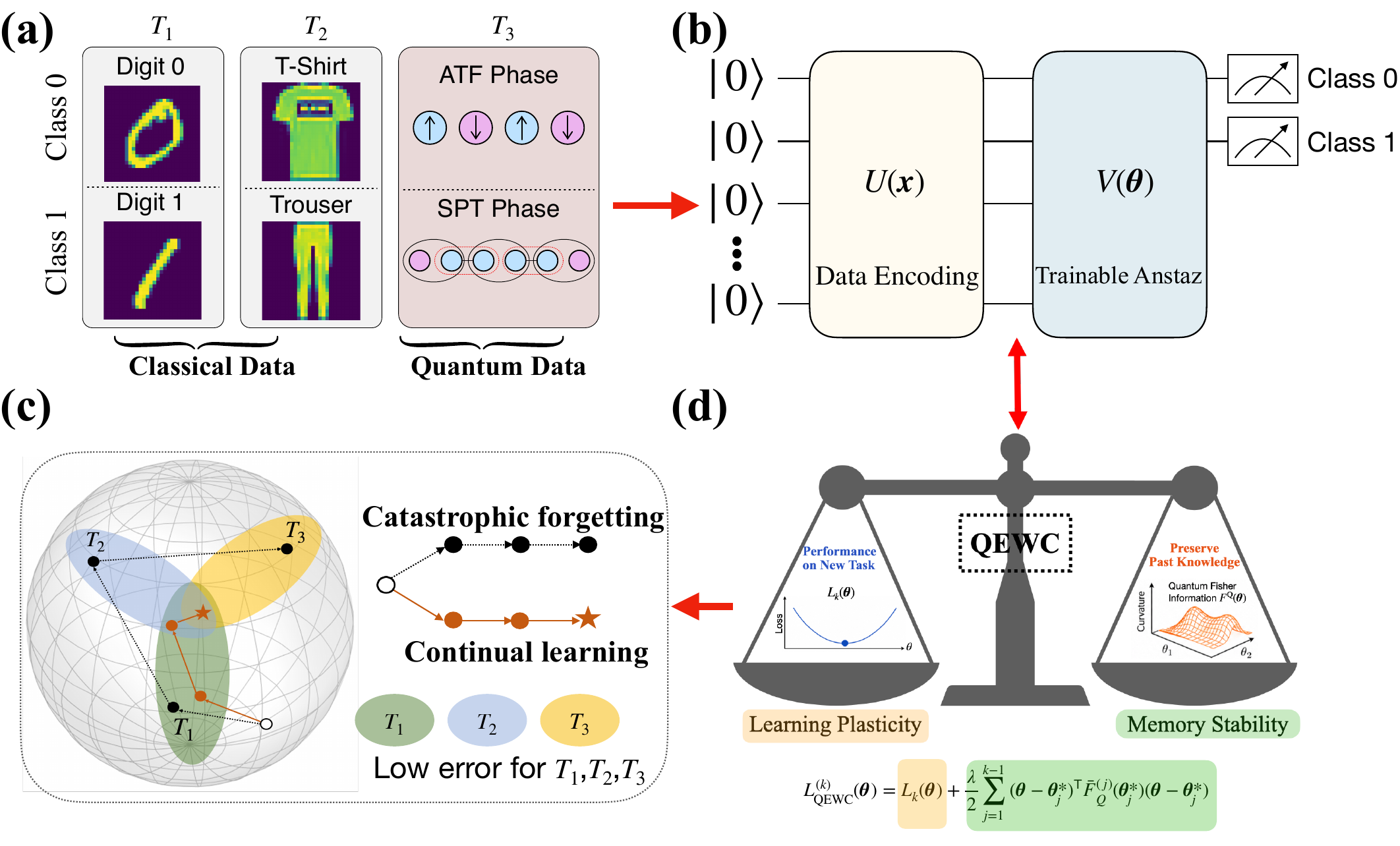}
\caption{
\textbf{Overview of the proposed quantum elastic weight consolidation (QEWC) framework.}
(a) continual learning tasks considered in this work, including classical image-classification tasks and a quantum-state classification task involving ATF and SPT phases.
(b) Variational quantum classifier used for sequential binary classification. Classical input data are embedded into the amplitudes of an $n$-qubit quantum state through amplitude encoding $U(\mathbf{x})$, processed by a trainable ansatz $V(\boldsymbol{\theta})$, and mapped to class predictions through measurement readout.
(c) Illustration of catastrophic forgetting and continual learning in the parameter/state manifold. Without knowledge consolidation, training on a new task can move the model away from regions associated with previous tasks, leading to performance degradation. Continual learning aims to maintain low error across all learned tasks.
(d) Schematic representation of the QEWC framework as a stability--plasticity balancing mechanism. The current-task loss promotes learning plasticity, while the QFI-informed regularization term penalizes parameter displacements that strongly affect previously learned quantum states, thereby preserving past knowledge.
}
\label{fig:QEWC}
\end{figure}

As discussed in the previous subsection, the CFI associated with a parametrized quantum state depends explicitly on the measurement statistics and, consequently, on the particular POVM employed. In the context of QNN, such a measurement-dependent quantity may not faithfully capture the intrinsic sensitivity of the quantum state to variations in the trainable parameters. This observation naturally motivates the use of the QFI, which provides a measurement-independent characterization of parameter importance in quantum continual learning.

We begin with the case in which the model has been trained on the first task $T_1$. Let $\boldsymbol{\theta}_1^{*}$ denote the corresponding optimal parameter vector. Evaluating the QFI matrix at this point, as given by Eq.~\eqref{eq:qfi_pure_full}, provides a local information-geometric characterization of how sensitively the quantum state responds to infinitesimal perturbations around the task optimum. The full QFI matrix captures both parameter-wise sensitivities and correlations between different parameter directions.

For the practical implementation of the QEWC regularizer, we adopt a diagonal approximation of the task-averaged QFI matrix. This approximation is consistent with the parameter-wise structure of the EWC framework and avoids the computational cost of estimating all off-diagonal parameter correlations.
Under this approximation, the importance weight assigned to parameter $\theta_i$ is given by the diagonal entry
$\left[\bar{F}_{Q}^{(1)}(\boldsymbol{\theta}_1^\ast)\right]_{ii}$,
which quantifies the average sensitivity of the quantum states generated for task $T_1$ to perturbations along the parameter direction $\theta_i$. Based on this parameter-wise information-geometric measure, the diagonal-QFI QEWC loss for learning a subsequent task $T_2$ is defined as
\begin{equation}
\mathcal{L}_{\mathrm{QEWC,diag}}
=
\mathcal{L}_{T_2}(\boldsymbol{\theta})
+
\frac{\lambda}{2}
\sum_i
\left[
\bar{F}_{Q}^{(1)}(\boldsymbol{\theta}_1^\ast)
\right]_{ii}
(\theta_i-\theta_{1,i}^{*})^2 ,
\label{eq:qewc_diagonal_two_task}
\end{equation}
where $\lambda$ is a hyperparameter controlling the regularization strength. The regularization term penalizes updates along parameter directions to which the previously learned quantum state is highly sensitive, while allowing comparatively larger changes along less sensitive directions. In this way, the QEWC framework preserves past-task knowledge by constraining parameter variations according to the local geometry of the quantum-state manifold.

More generally, let $\bar{F}_{Q}^{(j)}(\boldsymbol{\theta}_j^\ast)$ denote the task-averaged QFI matrix evaluated at the parameter vector obtained after training task $T_j$, as defined in Eq.~\eqref{eq:task_averaged_qfi}. When learning task $T_k$, the full-matrix QEWC objective is defined as

\begin{equation}
\mathcal{L}_{\mathrm{QEWC}}^{(k)}(\boldsymbol{\theta})
=
\mathcal{L}_{k}(\boldsymbol{\theta})
+
\frac{\lambda}{2}
\sum_{j=1}^{k-1}
(\boldsymbol{\theta}-\boldsymbol{\theta}_{j}^{*})^{\mathsf T}
\bar{F}_{Q}^{(j)}(\boldsymbol{\theta}_j^\ast)
(\boldsymbol{\theta}-\boldsymbol{\theta}_{j}^{*}) .
\label{eq:qewc_general}
\end{equation}

This objective penalizes parameter displacements according to the task-averaged QFI-induced quadratic form. Therefore, $\bar{F}_{Q}^{(j)}(\boldsymbol{\theta}_j^\ast)$ acts as the consolidation metric for the previously learned task $T_j$. When this matrix is positive definite, the corresponding quadratic form defines a local Riemannian metric~\cite{Braunstein1994} on the parametrized quantum-state manifold averaged over the empirical data distribution of task $T_j$. When the matrix is singular or rank deficient, the penalty acts only along parameter directions with nonzero QFI weight.
Although Eq.~\eqref{eq:qewc_general} gives the full geometric form of QEWC, estimating and storing the full task-averaged QFI matrix can be costly for VQC with many trainable parameters, since it requires access to pairwise correlations between all parameter directions. Therefore, in the numerical experiments, we use the diagonal surrogate of $\bar{F}_{Q}^{(j)}(\boldsymbol{\theta}_j^\ast)$, retaining only the parameter-wise sensitivities
$\left[\bar{F}_{Q}^{(j)}(\boldsymbol{\theta}_j^\ast)\right]_{ii}$
and neglecting the off-diagonal correlations. This leads to the practical $k$-task diagonal-QFI QEWC objective
\begin{equation}
\mathcal{L}_{\mathrm{QEWC,diag}}^{(k)}(\boldsymbol{\theta})
=
\mathcal{L}_{k}(\boldsymbol{\theta})
+
\frac{\lambda}{2}
\sum_{j=1}^{k-1}
\sum_i
\left[
\bar{F}_{Q}^{(j)}(\boldsymbol{\theta}_j^\ast)
\right]_{ii}
(\theta_i-\theta_{j,i}^{*})^2 .
\label{eq:qewc_diagonal}
\end{equation}

Eq.~\eqref{eq:qewc_diagonal_two_task} is recovered as the two-task special case of Eq.~\eqref{eq:qewc_diagonal}.
The essential distinction between the EWC and QEWC frameworks is therefore not merely the replacement of one Fisher matrix with another, but the choice of geometric object used to define parameter importance. In EWC, parameter importance is inferred from the sensitivity of a measurement-induced classical probability model. In QEWC, by contrast, parameter importance is determined by how strongly parameter variations change the underlying quantum states, averaged over the empirical data distribution of each previously learned task. In this sense, QEWC regularizes the geometry of the quantum-state manifold rather than only the geometry induced by a chosen classical readout distribution.

\subsection{Training and Optimization}

To investigate continual  learning in the presence of catastrophic forgetting, we consider two training frameworks based on the objective functions defined in Eq.~(\ref{eq:ewc_general_1}) and Eq.~(\ref{eq:qewc_general}), corresponding to the EWC and QEWC frameworks, respectively. 
In both cases, the quantum classifier is trained sequentially over the task stream $\{T_1,T_2,\ldots\}$ by minimizing the corresponding objective function at each task.

For task $T_k$, the trainable parameters are updated by minimizing a framework-dependent objective function $\mathcal{J}^{(k)}(\boldsymbol{\theta})$ using a gradient-based optimizer \footnote{In the numerical experiments, QEWC is implemented using the diagonal form of the task-averaged QFI matrix. The motivation for this choice and its comparison with the full-QFI implementation are discussed in ~\ref{app:full}}. 
More explicitly, we define
\begin{equation}
\mathcal{J}^{(k)}(\boldsymbol{\theta})
=
\begin{cases}
\mathcal{L}_{\mathrm{EWC}}^{(k)}(\boldsymbol{\theta}), & \text{for the EWC framework},\\
\mathcal{L}_{\mathrm{QEWC}}^{(k)}(\boldsymbol{\theta}), & \text{for the QEWC framework}.
\end{cases}
\label{eq:framework_objective}
\end{equation}

Denoting the optimization step by $t$, the parameter update rule is given by
\begin{equation}
\boldsymbol{\theta}^{(t+1)}
=
\boldsymbol{\theta}^{(t)}
-
\eta \,
\nabla_{\boldsymbol{\theta}}
\mathcal{J}^{(k)}(\boldsymbol{\theta}^{(t)}),
\label{eq:training_update}
\end{equation}
where $\eta$ denotes the learning rate.

% \en{we have use weight sharing, is the formula  (24) still valid?}

Since the quantum classifier is implemented as a PQC, the derivatives of the readout expectation values can be evaluated analytically via the parameter-shift rule~\cite{Schuld2019Evaluating,MitaraiQuantum2018,Wierichs2022generalparameter,Mari2021Estimating}. 
Using the two-qubit readout defined in Eq.~\eqref{eq:two_qubit_readout}, the derivative of each readout component with respect to the trainable parameter $\theta_i$ is given by
\begin{equation}
\frac{\partial z_c(\mathbf{x},\boldsymbol{\theta})}{\partial \theta_i}
=
\frac{1}{2}
\left[
z_c(\mathbf{x},\boldsymbol{\theta}_i^+)
-
z_c(\mathbf{x},\boldsymbol{\theta}_i^-)
\right],
\qquad c\in\{0,1\},
\label{eq:parameter_shift_readout}
\end{equation}
where $\boldsymbol{\theta}_i^\pm$ is obtained by shifting only the $i$-th component of $\boldsymbol{\theta}$, namely
$
\boldsymbol{\theta}_i^\pm
=
(\theta_0,\ldots,\theta_i\pm\pi/2,\ldots)^{\mathsf T}.
$
The gradient of the binary cross-entropy loss is then obtained by applying the chain rule through the softmax probability map in Eq.~\eqref{eq:softmax_readout}.

For the QEWC framework, the regularization term is differentiable in closed form. 
In particular, for the generalized QEWC objective in Eq.~\eqref{eq:qewc_general}, the derivative of the regularization term with respect to $\theta_i$ is given by

\begin{equation}
\frac{\partial}{\partial \theta_i}
\left[
\frac{\lambda}{2}
\sum_{j=1}^{k-1}
(\boldsymbol{\theta}-\boldsymbol{\theta}_{j}^{*})^{\mathsf T}
\bar{F}_{Q}^{(j)}(\boldsymbol{\theta}_j^{*})
(\boldsymbol{\theta}-\boldsymbol{\theta}_{j}^{*})
\right]
=
\lambda
\sum_{j=1}^{k-1}
\Big[
\bar{F}_{Q}^{(j)}(\boldsymbol{\theta}_j^{*})
(\boldsymbol{\theta}-\boldsymbol{\theta}_{j}^{*})
\Big]_i,
\label{eq:qewc_reg_grad}
\end{equation}

where $[\cdot]_i$ denotes the $i$-th component of the corresponding vector.

By combining the parameter-shift evaluation of the task-dependent loss with the analytic derivative of the QEWC regularizer, the full derivative of the generalized QEWC objective with respect to $\theta_i$ is obtained as

\begin{equation}
\frac{\partial \mathcal{L}_{\mathrm{QEWC}}^{(k)}(\boldsymbol{\theta})}{\partial \theta_i}
=
\frac{\partial \mathcal{L}_{k}(\boldsymbol{\theta})}{\partial \theta_i}
+
\lambda
\sum_{j=1}^{k-1}
\Big[
\bar{F}_{Q}^{(j)}(\boldsymbol{\theta}_j^{*})
(\boldsymbol{\theta}-\boldsymbol{\theta}_{j}^{*})
\Big]_i.
\label{eq:qewc_grad_general}
\end{equation}

The first term is evaluated from quantum circuit measurements using the parameter-shift rule~\cite{Schuld2019Evaluating,MitaraiQuantum2018,Wierichs2022generalparameter,Mari2021Estimating}, whereas the second term is computed analytically from the QEWC regularization term. 
The corresponding gradient for the EWC framework is obtained analogously by replacing the task-averaged QFI matrix $\bar{F}_{Q}^{(j)}(\boldsymbol{\theta}_j^{*})$ with the corresponding CFI matrix $F^{\mathrm C}(\boldsymbol{\theta}_j^{*})$.
Consequently, in both training frameworks, the optimization combines measurement-based gradient evaluation for the task-dependent loss with analytic updates induced by the regularization term. 
This hybrid gradient-evaluation strategy provides an efficient route for optimizing the model in sequential quantum learning.

\subsection{Theoretical Justification of QFI-Based Consolidation}
\label{sec:qfi_theoretical_justification}

We now provide a local theoretical justification for using the QFI as the consolidation metric in quantum continual learning. The central idea is that the QFI is not merely a positive semidefinite matrix used to weight a quadratic penalty. Rather, it is the local information metric that quantifies the distinguishability of neighboring quantum states. Therefore, suppressing the QFI-weighted parameter displacement locally suppresses the drift of the quantum states generated by the classifier. Since the classifier predictions are obtained from fixed readout observables followed by a softmax map, this state-level stability also limits the drift of the output probabilities and, under a mild nonzero-probability condition, controls the increase of the previous-task loss.

Consider a previously learned task $T_j$ with dataset $\mathcal{D}_j$. For each input $\mathbf{x}$, let $\rho_{\mathbf{x}}(\boldsymbol{\theta})$ denote the quantum state produced by the QNNs before the final measurement. In the noiseless setting, $\rho_{\mathbf{x}}(\boldsymbol{\theta})=|\psi(\mathbf{x},\boldsymbol{\theta})\rangle\langle\psi(\mathbf{x},\boldsymbol{\theta})|$, whereas in the noisy setting $\rho_{\mathbf{x}}(\boldsymbol{\theta})$ denotes the corresponding mixed state generated by the noisy circuit. Because the generated quantum state depends on the input, the relevant consolidation metric is the task-averaged QFI matrix $\bar{F}_{Q}^{(j)}(\boldsymbol{\theta}_j^\ast)$ defined in Eq.~\eqref{eq:task_averaged_qfi}. This matrix captures the average state sensitivity over the empirical data distribution of the previously learned task. In practice, the diagonal QEWC regularizer uses the diagonal entries $\left[\bar{F}_{Q}^{(j)}(\boldsymbol{\theta}_j^\ast)\right]_{ii}$ as parameter-wise importance weights.

The local effect of this QFI-weighted penalty follows directly from the relation between QFI and the Bures metric. 
For a sufficiently small displacement $\Delta\boldsymbol{\theta}=\boldsymbol{\theta}-\boldsymbol{\theta}_j^\ast$, the average squared Bures distance between the quantum states before and after the parameter update satisfies
\begin{equation}
\frac{1}{|\mathcal{D}_j|}
\sum_{(\mathbf{x},y)\in\mathcal{D}_j}
D_{\mathrm{Bures}}^2
\left(
\rho_{\mathbf{x}}(\boldsymbol{\theta}_j^\ast),
\rho_{\mathbf{x}}(\boldsymbol{\theta}_j^\ast+\Delta\boldsymbol{\theta})
\right)
=
\frac{1}{4}
\Delta\boldsymbol{\theta}^{\mathsf{T}}
\bar{F}_{Q}^{(j)}(\boldsymbol{\theta}_j^\ast)
\Delta\boldsymbol{\theta}
+
o\!\left(\|\Delta\boldsymbol{\theta}\|^2\right),
\label{eq:qfi_bures_local_expansion}
\end{equation}
where we use the convention that the QFI is four times the local Bures metric. 
Thus, controlling the QFI-weighted displacement locally controls the state-space drift induced by parameter updates.

This state-level control also constrains the classifier outputs because the readout observables are fixed. 
In our VQC, the class scores are computed from the expectation values of fixed Pauli observables,
$z_c(\mathbf{x},\boldsymbol{\theta})
=
\mathrm{Tr}\!\left[
\hat Z_c \rho_{\mathbf{x}}(\boldsymbol{\theta})
\right]$ for $c\in\{0,1\}$. 
The class probabilities are then obtained by applying the softmax map to the score vector $\mathbf{z}(\mathbf{x},\boldsymbol{\theta})$, namely
$p_{\boldsymbol{\theta}}(c|\mathbf{x})
=
\mathrm{softmax}\!\left(
\mathbf{z}(\mathbf{x},\boldsymbol{\theta})
\right)_c$. 
Since the Pauli observables are bounded and the softmax map is Lipschitz continuous, a small state-level drift induces only a controlled drift in the output probabilities. 
Consequently, the induced probability drift obeys
\begin{equation}
\frac{1}{|\mathcal{D}_j|}
\sum_{(\mathbf{x},y)\in\mathcal{D}_j}
\sum_c
\left|
p_{\boldsymbol{\theta}_j^\ast+\Delta\boldsymbol{\theta}}(c|\mathbf{x})
-
p_{\boldsymbol{\theta}_j^\ast}(c|\mathbf{x})
\right|
\le
O
\left(
\sqrt{
\Delta\boldsymbol{\theta}^{\mathsf{T}}
\bar{F}_{Q}^{(j)}(\boldsymbol{\theta}_j^\ast)
\Delta\boldsymbol{\theta}
}
\right)
+
o\!\left(\|\Delta\boldsymbol{\theta}\|\right).
\label{eq:qfi_probability_drift_bound}
\end{equation}
Furthermore, if the correct-label probabilities at the previous-task optimum are bounded away from zero, the same probability-drift control implies a local bound on the increase of the previous-task cross-entropy loss:
\begin{equation}
\mathcal{L}_j
\left(
\boldsymbol{\theta}_j^\ast+\Delta\boldsymbol{\theta}
\right)
-
\mathcal{L}_j
\left(
\boldsymbol{\theta}_j^\ast
\right)
\le
C_j
\sqrt{
\Delta\boldsymbol{\theta}^{\mathsf{T}}
\bar{F}_{Q}^{(j)}(\boldsymbol{\theta}_j^\ast)
\Delta\boldsymbol{\theta}
}
+
o\!\left(\|\Delta\boldsymbol{\theta}\|\right),
\label{eq:qfi_bce_local_bound}
\end{equation}
where $C_j>0$ is a task-dependent constant. 
The detailed proof of Eqs.~\eqref{eq:qfi_bures_local_expansion}--\eqref{eq:qfi_bce_local_bound} is provided in~\ref{app:qewc-bound}

The local bounds above clarify the role of the QFI in QEWC. They do not imply a global guarantee of preserving previous-task accuracy; rather, they show that the QFI provides a local stability metric for the quantum states generated by the classifier. When the parameter displacement is small in the QFI metric, the induced state drift is locally controlled. Because the readout observables are fixed and the softmax map is smooth, this state-level control also bounds the corresponding changes in readout scores, output probabilities, and the previous-task cross-entropy loss. Therefore, QEWC is justified as an information-geometric regularizer that discourages parameter updates causing large local changes in the quantum states associated with previously learned tasks.

The bounds are stated for the full task-averaged QFI matrix. In the numerical experiments, we use its diagonal approximation for computational efficiency. This diagonal surrogate retains parameter-wise sensitivities while neglecting correlations between different parameter directions. It should therefore be viewed as a practical coordinate-wise approximation to the full QFI metric, rather than as a fully coordinate-invariant geometric penalty.

\section{Experiment Setup}
We construct a comprehensive continual learning environment to systematically evaluate the effectiveness of the proposed QEWC framework in mitigating catastrophic forgetting. The environment comprises four binary classification tasks, denoted as $T_1$, $T_1'$, $T_2$, and $T_3$, and is designed to cover varying degrees of domain shift, ranging from structurally similar tasks to fundamentally distinct learning domains. Specifically, $T_1$ focuses on distinguishing between the handwritten digits ``0'' and ``1'' sampled from the MNIST dataset~\cite{MNIST1998}. $T_1'$ reuses the exact images from $T_1$ but applies a predefined fixed pixel permutation, thereby introducing a controlled distributional shift while preserving the underlying classification structure. In contrast, $T_2$ moves to the Fashion-MNIST dataset~\cite{Fashion2017}, requiring the classifier to distinguish between ``T-shirts'' and ``trousers''. Finally, $T_3$ constitutes a quantum-native problem, in which the objective is to classify many-body quantum states into either the antiferromagnetic (ATF) or the symmetry-protected topological (SPT) phase~\cite{Phase2011}.

We further investigate three progressive continual learning scenarios to examine the effect of increasing task heterogeneity: $T_1 \rightarrow T_1'$, $T_1 \rightarrow T_2$, and $T_1 \rightarrow T_2 \rightarrow T_3$. Each experiment is independently repeated across six trials with distinct random seeds to ensure statistical reliability. All reported quantities represent the mean and standard deviation computed over these six independent runs.

\subsection{Dataset Preparation and Encoding}
Regarding the classical image-classification tasks, namely $T_1$, $T_1'$, and $T_2$, the raw images are projected into sixteen-dimensional feature vectors using principal component analysis (PCA). For each sequence, the PCA transformation is fitted using the data from the first task and then kept fixed when transforming the subsequent task. These reduced classical features are then embedded into quantum states via amplitude encoding on a $4$-qubits register. For each task, the dataset contains $800$ training samples and $200$ testing samples.
In contrast, $T_3$ is a quantum-native task whose dataset consists of the exact ground states of the cluster-Ising Hamiltonian, which are numerically obtained through exact diagonalization. These ground-state vectors are directly supplied to the quantum classifier through amplitude encoding, which ensures that the topological signatures of the ATF and SPT phases are not affected by state-preparation infidelities. Further details regarding the quantum state preparation are provided in~\ref{app:data}

\subsection{Quantum Model Architecture and Training Setup}\label{sec:exp}
All numerical experiments are performed through classical simulation using the VQC architecture described in Sec.~\ref{sec:VQC}. In our implementation, the quantum classifier is realized on a $4$-qubit circuit with $30$ repeated variational layers. We further employ a weight-sharing mechanism, where the $R_y$ and $R_z$ rotation angles acting on the same qubit within the same layer are controlled by a shared trainable parameter. Although the circuit contains $240$ parametrized rotation gates in total, this constraint reduces the number of independent trainable parameters to $120$. Class-dependent scores are obtained by measuring the expectation values of Pauli-$Z$ observables on the first two qubits, and all gradients are evaluated analytically using the parameter-shift rule~\cite{Schuld2019Evaluating,MitaraiQuantum2018,Wierichs2022generalparameter,Mari2021Estimating}.

Within the EWC framework, the diagonal entries of the CFI matrix are computed using the empirical Fisher estimator, obtained from the squared log-likelihood gradients over the training dataset. In contrast, the diagonal entries of the QFI matrix in the QEWC framework are estimated using the diagonal approximation of the quantum geometric tensor defined in Eq.~\eqref{eq:qfi_pure_diag}, evaluated on the full circuit output state. The resulting QFI estimates are averaged over mini-batches of training data, yielding a dataset-averaged importance weight for each parameter. This diagonal approximation makes QFI estimation tractable for the $120$-parameter circuit considered here, although it may underestimate the full QFI when parameter correlations become significant, for example, in highly entangled circuit regimes.

To prevent the consolidation objective from assigning equal strength to all previously learned tasks, we employ a task-dependent weighting rule in the experiments. When learning the $k$-th task, the coefficient assigned to a previously learned task $T_j$ is defined as

\begin{equation}
\alpha_j^{(k)}=
\begin{cases}
\frac{j}{k-1}, & k > 1 \\
1, & k=1
\end{cases}
\label{eq:weight}
\end{equation}

This weighting schedule assigns smaller coefficients to tasks learned earlier in the sequence and larger coefficients to more recently learned tasks. For instance, when learning $T_3$, the weights are $\alpha_1^{(3)}=1/2$ and $\alpha_2^{(3)}=1$, so that the consolidation term associated with $T_1$ is weakened whereas the term associated with $T_2$ remains unchanged. Accordingly, the diagonal-QFI QEWC objective used in the numerical experiments is

\begin{equation}
\mathcal{L}_{\mathrm{QEWC,diag}}^{(k)}(\boldsymbol{\theta})
=
\mathcal{L}_{k}(\boldsymbol{\theta})
+
\frac{\lambda}{2}
\sum_{j=1}^{k-1}
\alpha_j^{(k)}
\sum_i
\left[
\bar{F}_{Q}^{(j)}(\boldsymbol{\theta}_j^\ast)
\right]_{ii}
(\theta_i-\theta_{j,i}^{*})^2 .
\label{eq:qewc_diagonal_weight}
\end{equation}

For a consistent comparison, the same task-dependent weighting rule is applied to the CFI-based EWC framework by replacing the task-averaged QFI matrix $\bar{F}^{(j)}_{\mathrm{Q}}$ in Eq.~\ref{eq:qewc_diagonal_weight} with the corresponding CFI matrix $F^{(j)}_{\mathrm{C}}$.

%\hsu{The coefficient $\alpha_j^{(k)}$ reweights the consolidation terms associated with previously learned tasks. 
%It weakens the regularization imposed by earlier tasks while maintaining stronger constraints from recently learned tasks.}

The VQC is optimized using the Adam optimizer~\cite{Adam2014} with a constant learning rate of $0.02$. Since the training set for each task contains $800$ samples, which is smaller than the batch size of $1024$, each optimization step effectively uses the entire training set and therefore corresponds to full-batch gradient descent. Each task in the continual-learning sequence is trained for $20$ epochs. To balance the stability--plasticity trade-off, the regularization strengths are empirically tuned and fixed at $\lambda=30$ for the CFI-based EWC approach and $\lambda_q=0.8$ for the QFI-based QEWC approach. The choice of these hyperparameters is critical because overly strong regularization can impose excessive parameter rigidity and hinder the acquisition of new knowledge, whereas overly weak regularization may insufficiently consolidate previously learned task representations.

\section{Results}

\subsection{Demonstration of Catastrophic Forgetting}\label{sec:forget}
\begin{figure}[!t]
\centering
\includegraphics[width=0.82\textwidth]{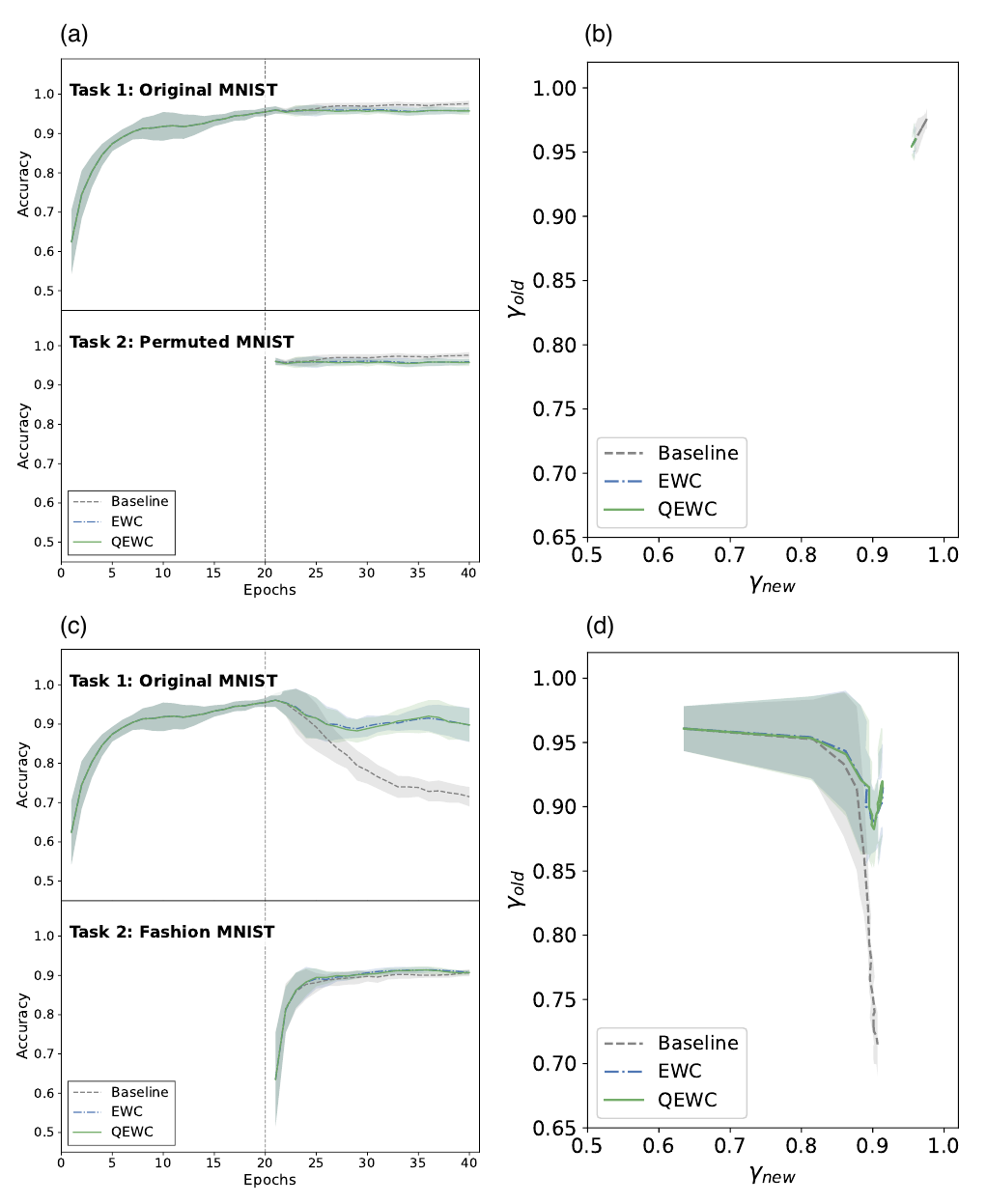}
\caption{\textbf{Performance benchmarking for the EWC and QEWC frameworks.} (a) Learning curves of two similar tasks: classifying the original and pixel-permuted MNIST images. Gray lines plot the accuracies for the two tasks respectively trained without the EWC or QEWC framework, whereas blue and green lines show the corresponding results with using of EWC and QEWC frameworks respectively. (b) Forgetting curve for two tasks with large similarity. Here, $\gamma_{\text{old}}$ ($\gamma_{\text{new}}$) represents the accuracy of the quantum classifier on the old (new) task. (c) Learning curves of two dissimilar tasks: classifying the original MNIST hand-written images and fashion-MNIST clothing images. (d) Forgetting curve of two dissimilar tasks: the classifications hand-written and digits and clothing images.}
\label{fig: task_12}
\end{figure}

\begin{figure}[!t]
\centering
\includegraphics[width=0.82\textwidth]{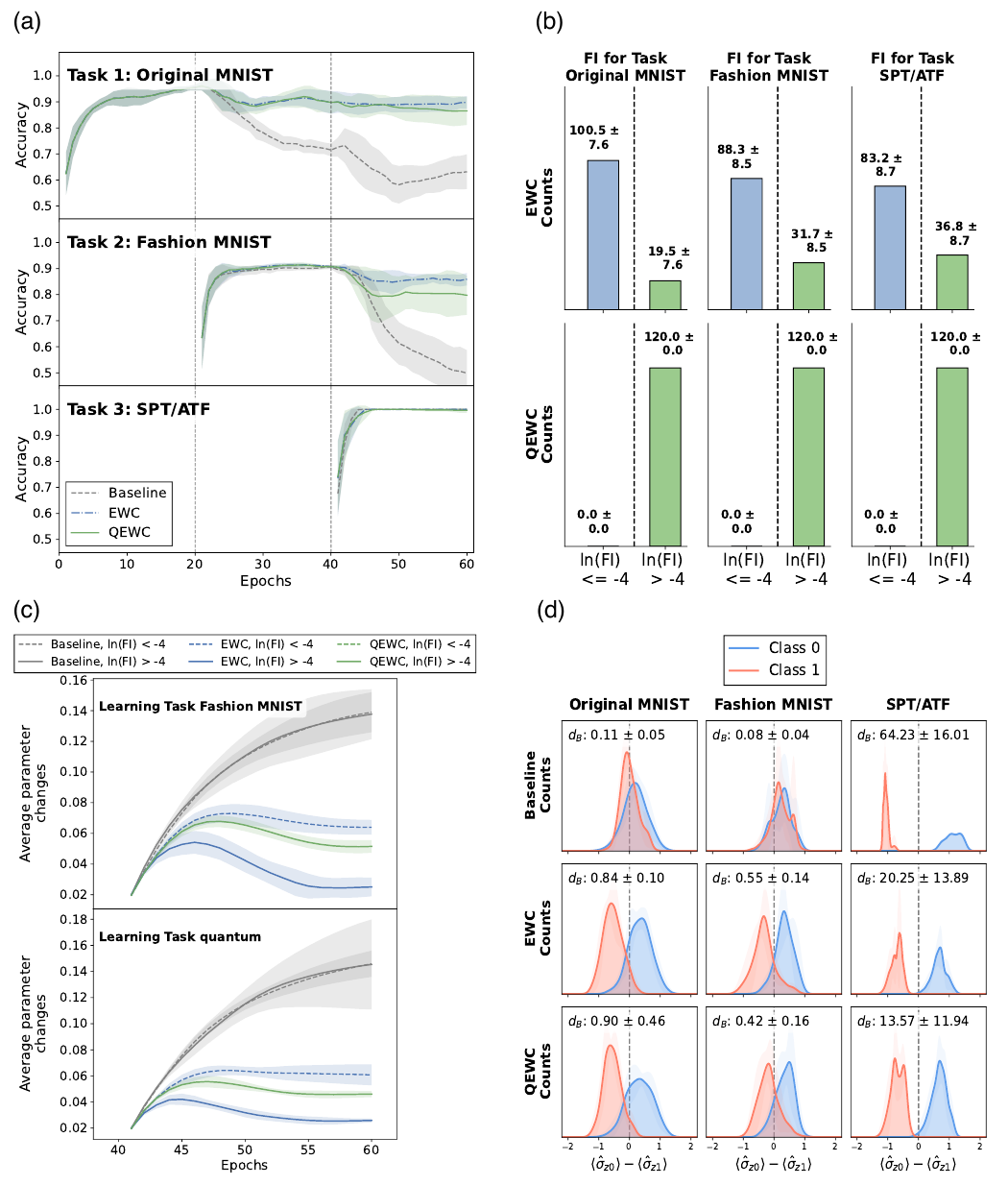}
\caption{\textbf{Experimental results for continually learning three tasks.} (a) The prediction accuracy for three sequential tasks at each epoch during the continual learning process of the quantum classifier. (b) Distribution of Fisher information for all parameters after learning each task. (c) Average parameter change compared to the obtained parameters for previous tasks during the learning stage for the new task. The top (bottom) figure corresponds to the learning for $T_2$ ($T_3$)  (d) Distribution of the measured expected values which determine the prediction label of input data. For each task, red and blue correspond to two classes of data samples, respectively. A greater separation between the two distributions means better classification performance}
\label{fig:task_134}
\end{figure}

To establish a systematic benchmark for analyzing catastrophic forgetting in sequential quantum learning, we first evaluate an unregularized baseline, namely a standard VQC trained without any knowledge-consolidation mechanism, across all three task sequences. The corresponding learning curves are shown in Figs.~\ref{fig: task_12} and~\ref{fig:task_134}.
We begin with the $T_1 \rightarrow T_1'$ sequence, in which the second task retains the original MNIST digit images but applies a fixed pixel permutation. This introduces a controlled distributional shift while preserving the underlying classification structure. As illustrated in Fig.~\ref{fig: task_12}a, the model achieves an accuracy of $95.5\%$ on $T_1$ at the task transition boundary. After the onset of $T_1'$ training, the $T_1$ accuracy does not deteriorate; instead, it continues to improve throughout the second training stage and converges to $97.6\%$ by the end of training, while the accuracy on $T_1'$ also reaches $97.6\%$. This behavior is consistent with the intuition that structurally similar tasks impose overlapping demands on the parameter space. Since both tasks require the classifier to distinguish between the same digit categories, differing only in their spatial arrangement, the optimal parameter configurations for $T_1$ and $T_1'$ exhibit a substantial degree of compatibility. As a result, training on $T_1'$ does not induce destructive interference with the knowledge encoded from $T_1$. Consequently, this regime does not exhibit genuine catastrophic forgetting and serves as a useful limiting case for the following analysis.

The $T_1 \rightarrow T_2$ sequence provides a more challenging test case, since the second task belongs to the qualitatively distinct Fashion-MNIST domain. 
As shown in Fig.~\ref{fig: task_12}c, the model reaches an accuracy of $90.6\%$ on $T_2$ by the end of the second training stage, indicating successful adaptation to the new task.
However, the performance on $T_1$ undergoes severe degradation, declining from $95.5\%$ at the task transition to $71.5\%$ after the completion of $T_2$ training, corresponding to a drop of approximately $24\%$ points.
The most severe forgetting is observed in the three-task sequence $T_1 \rightarrow T_2 \rightarrow T_3$, which further incorporates the quantum phase-classification task $T_3$. As shown in Fig.~\ref{fig:task_134}a, the model achieves a classification accuracy of $100\%$ on $T_3$ by the end of the final training stage. However, this strong performance is obtained at a substantial cost to previously acquired knowledge: the accuracy on $T_1$ decreases to $63.2\%$, corresponding to a reduction of approximately $32\%$ points from its peak value, while the accuracy on $T_2$ collapses to $49.8\%$, which is effectively indistinguishable from random guessing in binary classification.

This progressive deterioration of earlier tasks with each additional training stage directly reflects the compounding nature of catastrophic forgetting. Each successive optimization stage, being directed toward a newly encountered task, induces additional displacements away from the optima associated with preceding tasks and cumulatively erodes the knowledge structure encoded in the parameter vector. Furthermore, the degradation of $T_2$ after introducing $T_3$ is more pronounced than that of $T_1$, suggesting that the parameter configurations required for quantum phase classification are particularly incompatible with those needed for image-domain tasks. This behavior reflects the substantial domain mismatch between classical image classification and quantum-native learning. Taken together, these observations establish catastrophic forgetting as a central obstacle in quantum continual learning, with its severity governed by both the heterogeneity of the task sequence and the cumulative number of sequentially encountered tasks.

To establish a systematic benchmark for analyzing catastrophic forgetting in sequential quantum learning, we first evaluate an unregularized baseline, namely a standard VQC trained without any knowledge-consolidation mechanism, across all three task sequences. The corresponding learning curves are shown in Figs.~\ref{fig: task_12} and~\ref{fig:task_134}.
We begin with the $T_1 \rightarrow T_1'$ sequence, in which the second task retains the original MNIST digit images but applies a fixed pixel permutation. This introduces a controlled distributional shift while preserving the underlying classification structure. As illustrated in Fig.~\ref{fig: task_12}a, the model achieves an accuracy of $95.5\%$ on $T_1$ at the task transition boundary. After the onset of $T_1'$ training, the $T_1$ accuracy does not deteriorate; instead, it continues to improve throughout the second training stage and converges to $97.6\%$ by the end of training, while the accuracy on $T_1'$ also reaches $97.6\%$. This behavior is consistent with the intuition that structurally similar tasks impose overlapping demands on the parameter space. Since both tasks require the classifier to distinguish between the same digit categories, differing only in their spatial arrangement, the optimal parameter configurations for $T_1$ and $T_1'$ exhibit a substantial degree of compatibility. As a result, training on $T_1'$ does not induce destructive interference with the knowledge encoded from $T_1$. Consequently, this regime does not exhibit genuine catastrophic forgetting and serves as a useful limiting case for the following analysis.

The $T_1 \rightarrow T_2$ sequence provides a more challenging test case, since the second task belongs to the qualitatively distinct Fashion-MNIST domain. 
As shown in Fig.~\ref{fig: task_12}c, the model reaches an accuracy of $90.6\%$ on $T_2$ by the end of the second training stage, indicating successful adaptation to the new task.
However, the performance on $T_1$ undergoes severe degradation, declining from $95.5\%$ at the task transition to $71.5\%$ after the completion of $T_2$ training, corresponding to a drop of approximately $24\%$ points.
The most severe forgetting is observed in the three-task sequence $T_1 \rightarrow T_2 \rightarrow T_3$, which further incorporates the quantum phase-classification task $T_3$. As shown in Fig.~\ref{fig:task_134}a, the model achieves a classification accuracy of $100\%$ on $T_3$ by the end of the final training stage. However, this strong performance is obtained at a substantial cost to previously acquired knowledge: the accuracy on $T_1$ decreases to $63.2\%$, corresponding to a reduction of approximately $32\%$ points from its peak value, while the accuracy on $T_2$ collapses to $49.8\%$, which is effectively indistinguishable from random guessing in binary classification.

This progressive deterioration of earlier tasks with each additional training stage directly reflects the compounding nature of catastrophic forgetting. Each successive optimization stage, being directed toward a newly encountered task, induces additional displacements away from the optima associated with preceding tasks and cumulatively erodes the knowledge structure encoded in the parameter vector. Furthermore, the degradation of $T_2$ after introducing $T_3$ is more pronounced than that of $T_1$, suggesting that the parameter configurations required for quantum phase classification are particularly incompatible with those needed for image-domain tasks. This behavior reflects the substantial domain mismatch between classical image classification and quantum-native learning. Taken together, these observations establish catastrophic forgetting as a central obstacle in quantum continual learning, with its severity governed by both the heterogeneity of the task sequence and the cumulative number of sequentially encountered tasks.

\subsection{Continual Learning of the EWC and QEWC frameworks}\label{sec:CL}

Having characterized the extent of catastrophic forgetting in the unregularized baseline, we now evaluate the effectiveness of the EWC and QEWC frameworks in mitigating this phenomenon across the same task sequences. In both regularized frameworks, the VQC is trained with the modified loss functions defined in Eqs.~\eqref{eq:ewc_general_1} and~\eqref{eq:qewc_diagonal}, respectively.

In the $T_1 \rightarrow T_1'$ sequence, both methods successfully preserve the knowledge acquired from $T_1$ throughout the second training stage, as shown in Figs.~\ref{fig: task_12}a and~\ref{fig: task_12}b. After completing $T_1'$ training, the EWC framework attains an accuracy of $95.9\%$ on both tasks, while the QEWC framework achieves the accuracies of $95.6\%$ on both tasks. This result indicates that the consolidation penalty is not critically required when the two tasks share a substantial degree of structural similarity, consistent with the intuition that tasks with overlapping representations do not induce strong parameter competition.

For the $T_1 \rightarrow T_2$ sequence, both regularization approaches substantially mitigate the catastrophic forgetting observed in the unregularized baseline. As illustrated in Figs.~\ref{fig: task_12}c and~\ref{fig: task_12}d, the EWC framework preserves the $T_1$ accuracy at $89.6\%$, while the QEWC framework retains it at $89.8\%$ after the completion of $T_2$ training; both values represent clear improvements over the baseline accuracy of $71.5\%$. At the same time, both methods maintain strong performance on the newly learned task, with both the EWC and QEWC framework reaching a final $T_2$ accuracy of $90.8\%$. The two methods therefore exhibit comparable stability-plasticity behavior in this two-task sequence, with both frameworks preserving the previously learned task while maintaining high accuracy on the newly introduced task.

The stability--plasticity behavior becomes more subtle in the three-task sequence $T_1 \rightarrow T_2 \rightarrow T_3$. During the third learning phase, the task-dependent consolidation rule defined in Eq.~\ref{eq:weight} assigns coefficients $\alpha_1^{(3)}=1/2$ and $\alpha_2^{(3)}=1$ to the consolidation terms associated with $T_1$ and $T_2$, respectively. The corresponding accuracy dynamics are shown in Fig.~\ref{fig:task_134}a. Compared with the unregularized baseline discussed in Sec.~\ref{sec:forget}, both regularized frameworks retain substantially higher accuracies on the previously learned tasks while still learning the final quantum phase-classification task. The EWC framework reaches $100\%$ accuracy on $T_3$ and preserves accuracies of $89.8\%$ and $85.9\%$ on $T_1$ and $T_2$, respectively. The QEWC framework also maintains a high final-task accuracy of $99.6\%$, while retaining accuracies of $86.5\%$ and $79.8\%$ on $T_1$ and $T_2$, respectively.

We note that, during the third learning phase, the accuracy of QEWC on the intermediate task $T_2$ decreases more noticeably than that of EWC, although it remains well above the unregularized baseline. This behavior suggests that the CFI- and QFI-based penalties impose different constraints on the optimization trajectory. As discussed in Sec.~\ref{sec:mechanistic}, this difference can be understood from their distinct FI spectra and the resulting regularization geometries.
\section{Mechanistic Analysis of the EWC and the QEWC frameworks}\label{sec:mechanistic}

We further analyze how the EWC and QEWC frameworks mitigate catastrophic forgetting, we analyze the internal optimization dynamics of the quantum classifier using three complementary diagnostics. These diagnostics consist of the distribution of Fisher information (FI) values across parameter space, the average parameter displacement grouped by Fisher importance, and the statistical separability of the output distributions. Together, they provide a multi-scale view that connects the regularization geometry to parameter-level dynamics and, ultimately, to output-level task behavior.

The FI distribution first identifies which parameter directions are penalized and how strongly they are weighted by the regularizer.  However, the FI values alone do not determine whether these penalties effectively constrain the optimizer during training. 
We therefore analyze parameter displacements grouped by Fisher importance to test whether the imposed regularization geometry leads to differential suppression of parameter updates. Finally, because constrained parameter motion does not necessarily guarantee preservation of task-specific representations, we examine the output distributions to determine whether the resulting dynamics maintain or degrade class separability for each task. Taken together, these diagnostics clarify how the choice of FI metric shapes the optimization trajectory and consequently affects sequential task performance.

\subsection{Fisher Information Spectra}
\label{sec:fi_spectra}

As depicted in Fig.~\ref{fig:task_134}b, the EWC and QEWC frameworks exhibit markedly different FI spectral structures. 
Because the CFI and QFI values differ substantially in magnitude, we present the FI spectra on a logarithmic scale using $\ln(\mathrm{FI})$ for visualization. 
When discussing the magnitude of the regularization weights below, we report the corresponding values in the original FI scale.

For the EWC framework, the threshold $\ln(\mathrm{FI})>-4$, corresponding to $\mathrm{FI}\simeq 0.018$ in the original scale, separates a sparse set of high-CFI parameters from the remaining low-CFI directions. 
After learning $T_1$, approximately $19$ parameters exceed this threshold; this number increases to $31$ after learning $T_2$ and $36$ after learning $T_3$. 
The corresponding high-CFI values are typically of order $10^{-2}$, with mean values around $0.02$--$0.03$. 
These observations indicate that the CFI-based regularizer assigns relatively strong penalties to a limited subset of measurement-sensitive parameters, while leaving a large fraction of the parameter space only weakly constrained.

The QFI spectrum exhibits a qualitatively different structure from the CFI spectrum. 
Using the same threshold, all $120$ variational parameters are classified as high-QFI directions for every task. 
Moreover, the QFI values are substantially larger than the CFI values, typically lying around $0.4$--$0.5$ in the original scale. 
With this threshold, the QFI spectrum does not separate the parameters into sparse high- and low-FI groups in the same way as the CFI spectrum. 
Instead, QEWC imposes a dense state-geometric constraint over the parameter space. 
This distinction indicates that EWC acts as a more selective regularizer based on measurement-induced output sensitivity, whereas QEWC constrains parameter motion according to the broader geometric sensitivity of the parameterized quantum state.

The dense QFI spectrum may also help explain the more noticeable decrease in QEWC accuracy on the intermediate task $T_2$ during the third learning phase. 
Because almost all parameters receive relatively large QFI weights, QEWC constrains the parameter trajectory more globally than EWC. 
This global constraint improves retention relative to the unregularized baseline, but it offers less parameter-wise selectivity than the sparse CFI spectrum. 
This reduced selectivity provides a possible explanation for why QEWC retains $T_2$ less effectively than EWC in the noiseless three-task setting.

\subsection{Parameter Displacement Under Fisher Regularization}
\label{sec:param_dynamics}

The FI spectra characterize the regularization weights, but they do not directly show whether the optimizer is effectively constrained by it. To examine this point, we track the mean absolute displacement of parameter groups relative to the previous task optimum throughout each training stage, as shown in Fig.~\ref{fig:task_134}c. 

For the EWC framework, the threshold naturally separates the parameters into high-CFI and low-CFI groups. 
During subsequent training, the high-CFI group exhibits a smaller average displacement than the low-CFI group. 
This confirms that the CFI-based penalty effectively suppresses updates along the parameter directions identified as important for the previous task. 
In this sense, EWC mitigates forgetting through a selective stabilization mechanism in parameter space.

For the QEWC framework, the same threshold does not yield a meaningful separation between high-QFI and low-QFI groups, because all parameters lie above the threshold. 
Rather than selectively locking a sparse subset of parameters, QEWC acts as a global state-geometric constraint on the variational circuit. 
The relatively small parameter motion observed under QEWC indicates that the QFI-based penalty restricts updates over a broad range of state-sensitive directions. 
This behavior is consistent with the FI spectra discussed above and suggests that the two frameworks mitigate forgetting through different regularization geometries: CFI-based EWC selectively suppresses a subset of output-sensitive parameters, whereas QFI-based QEWC constrains the overall motion of the quantum-state manifold.

\subsection{Output Distribution Separability}
\label{sec:output_separability}

The downstream consequence of the learned parameter configuration is quantified using the Bhattacharyya distance~\cite{Bhattacharyya1946} ($d_B$) between the output distributions of the two classes for each task. 
The distance is evaluated after completing the training on $T_3$, and the results are shown in Fig.~\ref{fig:task_134}d. 
Specifically, $d_B$ measures the statistical separability between the class-conditional readout distributions obtained from the Pauli-$Z$ measurement operators, $\hat{Z}_0$ and $\hat{Z}_1$, acting on the two readout qubits, with larger values indicating more clearly separated class representations.

The output-distribution analysis further supports the accuracy trends reported in Sec.~\ref{sec:CL}. 
For the previously learned tasks $T_1$ and $T_2$, the unregularized baseline exhibits near-complete distributional collapse, with $d_B=0.11$ for $T_1$ and $d_B=0.08$ for $T_2$. 
Both regularized frameworks instead preserve clearer class separability for the earlier tasks. 
Specifically, EWC yields $d_B=0.84$ for $T_1$ and $d_B=0.55$ for $T_2$, while QEWC yields $d_B=0.90$ for $T_1$ and $d_B=0.42$ for $T_2$. 
These results indicate that Fisher-based regularization not only improves retention accuracy, but also helps maintain distinguishable output distributions for previously learned tasks.

For the final task $T_3$, all methods produce well-separated output distributions after training. The unregularized baseline gives the largest separability, with $d_B=64.23$, whereas EWC and QEWC yield $d_B=20.25$ and $d_B=13.57$, respectively. This large baseline separability on $T_3$ should not be interpreted as better continual learning. Instead, it reflects strong specialization to the final task, which occurs together with severe collapse of the earlier-task output distributions. In contrast, the regularized methods produce smaller final-task separability but preserve substantially better separability for $T_1$ and $T_2$. Therefore, the output-distribution results support the same conclusion as the accuracy curves: EWC and QEWC mitigate catastrophic forgetting by maintaining class-separable representations for previous tasks while still allowing the model to learn the final quantum phase classification task.

\section{Quantum noise}

\begin{figure*}[!t]
    \centering
    
    \begin{subfigure}[t]{0.48\textwidth}
        \centering
        \caption{}
        \vspace{-2pt}
        \label{fig:acc_noise}
        \includegraphics[width=\linewidth]{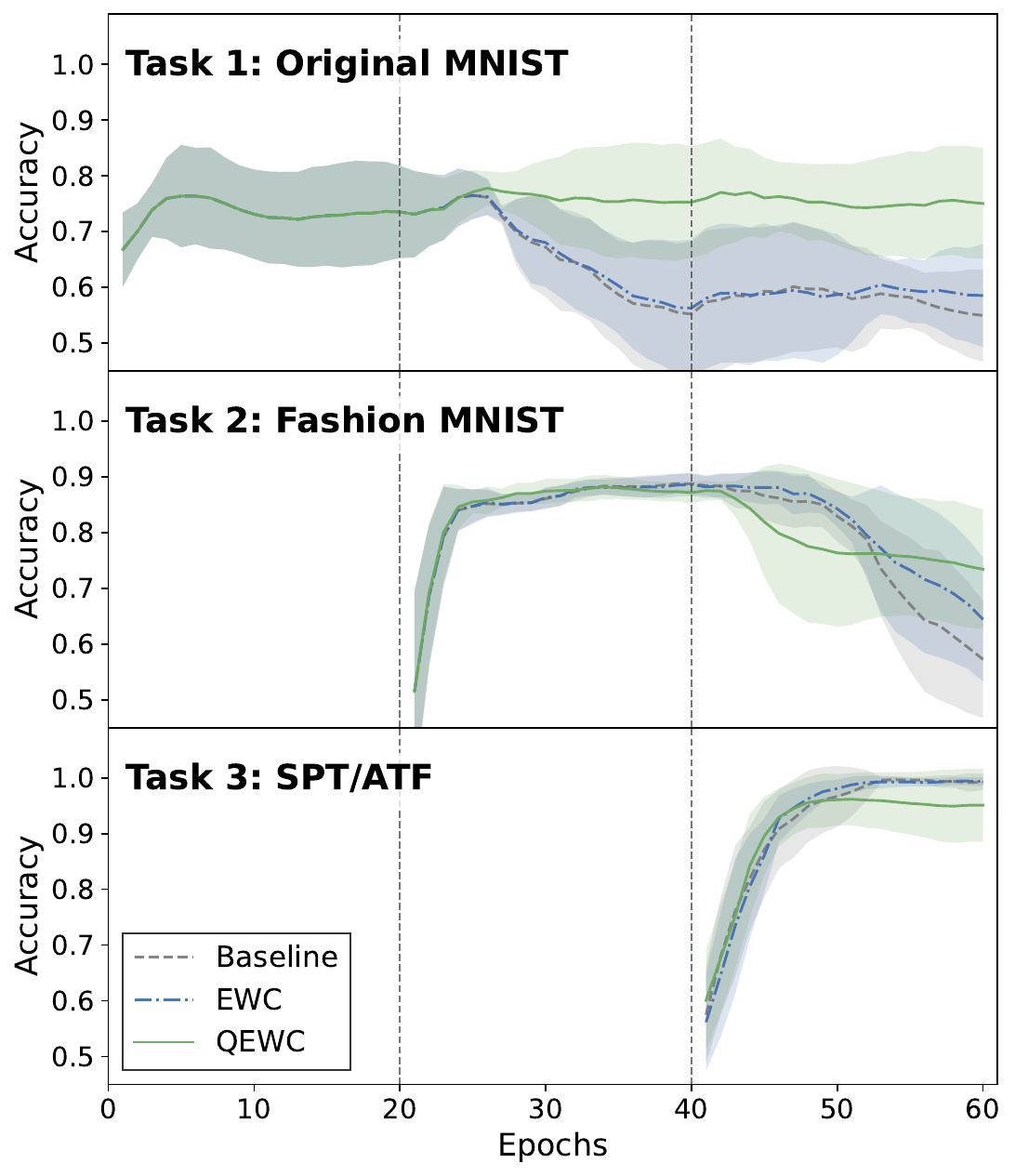}
    \end{subfigure}%
    \hfill
    \begin{subfigure}[t]{0.48\textwidth}
        \centering
        \caption{}
        \vspace{-2pt}
        \label{fig:param_change_noise_task_3}
        \includegraphics[width=\linewidth]{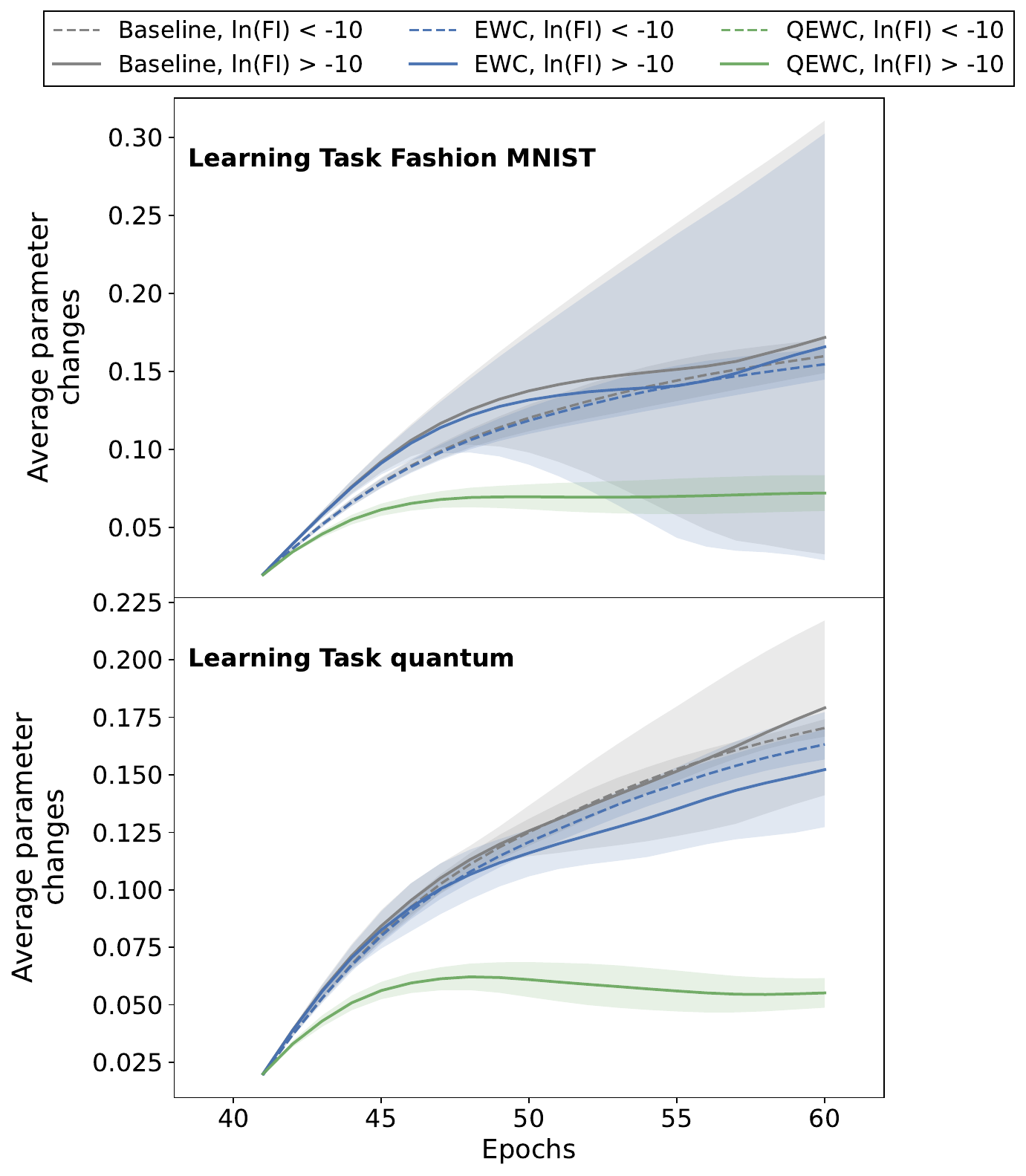}
    \end{subfigure}
    \vspace{-5pt}
    \caption{\textbf{Continual-learning performance and parameter dynamics under noise for the three-task sequence.} (a) Classification accuracy of the baseline, EWC, and QEWC frameworks at each epoch under depolarizing noise ($p=0.01$), with dashed vertical line indicating task boundaries. (b) Mean absolute parameter displacement of the high-FI and low-FI groups relative to the previous task optimum during $T_2$ and $T_3$ training. }
    \label{fig: task_noise_3}
\end{figure*}

\begin{figure*}[t]
\centering
\includegraphics[width=0.44\textwidth]{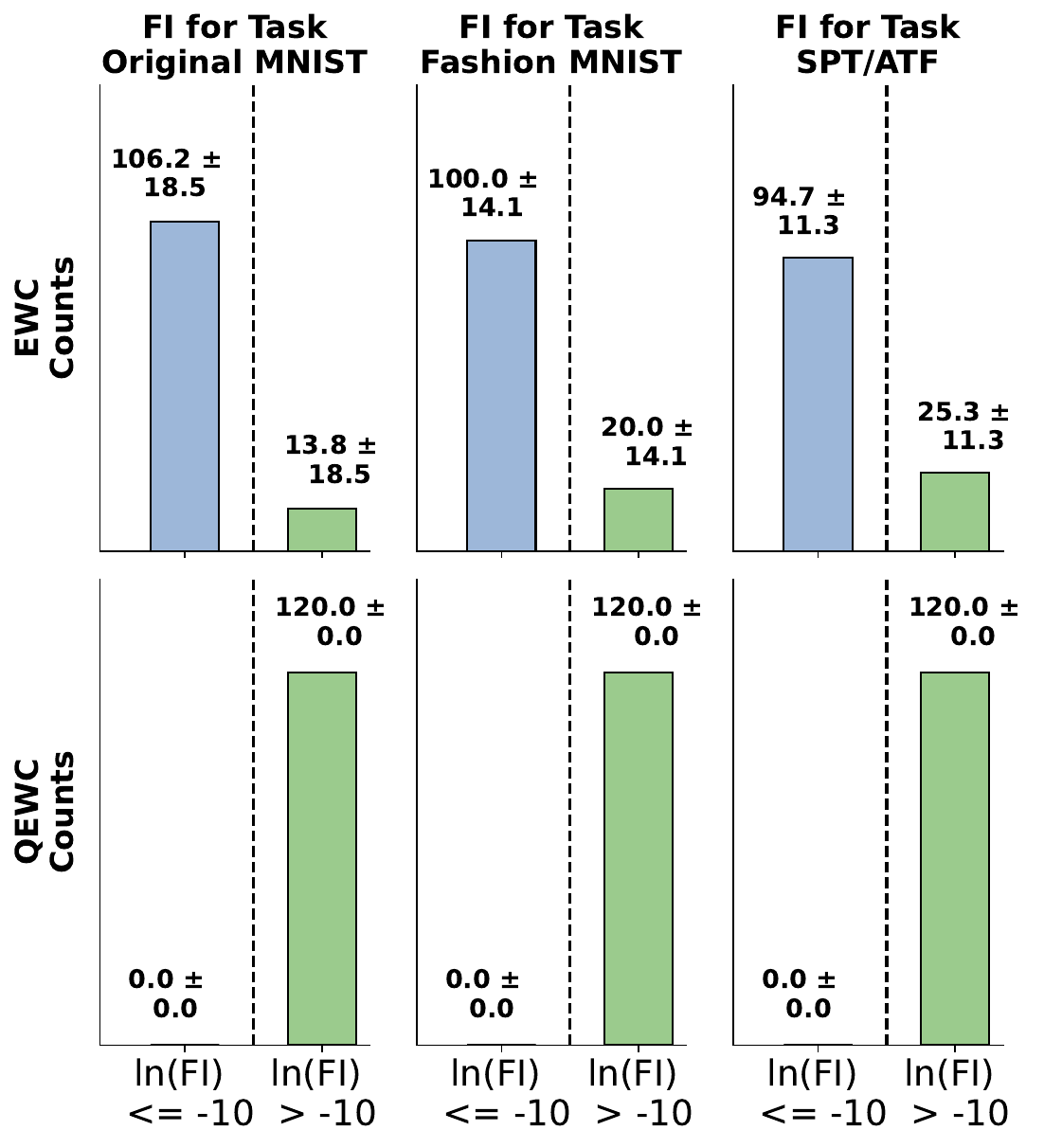}
\caption{\textbf{Fisher information spectra under depolarizing noise.} }
\label{fig: noise_counts}
\end{figure*}

To examine the robustness of the proposed method under more realistic NISQ conditions, we evaluate quantum continual learning in the presence of quantum noise. 
In contrast to the noiseless simulations considered above, where the VQC evolves pure quantum states, noisy evolution generally maps the classifier output to a mixed density operator. 
Consequently, the FI metrics used by the regularization frameworks must be evaluated from the mixed-state density matrix rather than from the pure-state expressions used in the previous sections.

We model gate noise using depolarizing channels~\cite{FanizzaDepolarizing2020,King2003depolarizing} inserted throughout the circuit. 
Specifically, the noise channel is applied after the amplitude-encoding stage, after each CNOT entangling operation on the involved qubits, and after every $R_y$ and $R_z$ rotation. 
A uniform depolarizing strength of $p=0.01$ is used for all noise channels, and all numerical simulations in this section are performed using a density-matrix simulator~\cite{bergholm2022}.

The CFI used in EWC is computed from the squared loss gradients over the training data, following the same procedure as in the noiseless setting but using the noisy circuit and the task-dependent weighting factor $\alpha_j^{(k)}$ introduced in Sec.~\ref{sec:CL}. 
For QEWC, the QFI matrix is evaluated using the mixed-state spectral representation in Eq.~\eqref{eq:QFI_spectral}, rather than the pure state Fubini--Study expression in Eq.~\eqref{eq:qfi_pure_full}. 
This distinction is important because noise suppresses coherences and redistributes populations among eigenmodes, thereby modifying both the scale and structure of the QFI weights. 
The mixed-state expression therefore captures both eigenvalue and eigenvector contributions to the geometric sensitivity of the density matrix $\rho(\boldsymbol{\theta})$. 
The QFI is then estimated using the diagonal approximation of the quantum geometric tensor evaluated on the noisy parameterized state, with depolarizing channels included in the circuit definition. 
Under noisy dynamics, the QEWC regularization strength is tuned to $\lambda_q=0.03$, which provides a suitable stability--plasticity balance in the noise-modified optimization landscape. 
For consistency, the EWC regularization strength is kept fixed at $\lambda=30$.

The accuracy dynamics under depolarizing noise are shown in Fig.~\ref{fig:acc_noise}. 
Compared with the noiseless results, all methods exhibit reduced overall accuracy, reflecting the additional optimization difficulty introduced by the noisy loss landscape. The unregularized baseline still learns the final task effectively, but suffers substantial forgetting on the earlier tasks.

The EWC framework shows only a modest improvement over the unregularized baseline, indicating that CFI-based consolidation becomes less effective when the measurement-induced output statistics are degraded by noise. By contrast, QEWC provides a more favorable stability--plasticity balance in the noisy setting: although its final-task accuracy is slightly lower than those of the baseline and EWC, it retains substantially higher accuracies on both $T_1$ and $T_2$ while still learning $T_3$ with high accuracy.

In order to understand this behavior, we analyze the FI spectra under noisy dynamics, as shown in Fig.~\ref{fig: noise_counts}. Because the overall FI scale is reduced in the noisy setting, we use the threshold $\ln(\mathrm{FI})=-10$, corresponding to $\mathrm{FI}\simeq 4.5\times10^{-5}$ in the original scale. For EWC, the CFI values are strongly suppressed compared with the noiseless case, with the mean high-CFI values remaining only of order $10^{-5}$. This suggests that noisy measurement statistics provide a weak and less stable parameter-importance signal for the EWC penalty. In contrast, the QFI values used by QEWC remain substantially larger in magnitude and preserve a more stable geometric sensitivity signal under mixed-state dynamics.

This difference is reflected in the parameter-displacement results shown in Fig.~\ref{fig:param_change_noise_task_3}. For EWC, the average parameter displacement remains close to that of the unregularized baseline, suggesting that the noise-suppressed CFI weights impose only a weak constraint on the optimizer. The high-CFI group also exhibits relatively large fluctuations, indicating less stable parameter-wise constraints across different runs. For QEWC, the average parameter displacement remains consistently smaller than those of both the baseline and EWC cases. Thus, the mixed-state QFI penalty continues to restrict parameter motion under depolarizing noise, which is consistent with the improved retention observed for the earlier tasks.

\section{Conclusion}

In this work, we introduce QEWC, a QFI-informed regularization framework for mitigating catastrophic forgetting in quantum continual learning. 
For comparative studies, we consider three training strategies, including an unregularized VQC baseline, the CFI-based EWC framework, and the proposed QFI-based QEWC framework. 
We apply these methods to sequential binary classification tasks and evaluate their performance in both noiseless and noisy simulation settings. 
Through a series of comparative analyses, we not only demonstrate that regularization-based approaches can substantially mitigate catastrophic forgetting in VQC, but also show that the QFI provides a distinct perspective on quantum continual learning by shifting the notion of parameter importance from measurement-dependent output statistics to the intrinsic geometry of parameterized quantum states.

Notably, continual learning in quantum models presents a distinct challenge because parameter importance is not merely a property of the classical output distribution, but is also tied to the geometry of the underlying quantum state generated by the parameterized circuit. 
The conventional EWC framework relies on the CFI, which depends explicitly on the chosen measurement statistics and therefore provides a measurement-dependent characterization of sensitivity. 
To address this limitation, we replace the CFI regularization metric with the QFI, which quantifies the local distinguishability of neighboring quantum states and provides a measurement-independent description of state sensitivity. 
We further provide a local theoretical justification showing that the QFI-weighted parameter displacement controls the drift of the quantum states generated by the classifier and, through fixed readout observables followed by the softmax map, the drift of the output probabilities and the previous-task cross-entropy loss.

Furthermore, our mechanistic analysis reveals that the EWC and QEWC frameworks mitigate forgetting through qualitatively different parameter-level dynamics. 
The CFI used by EWC identifies a comparatively sparse subset of measurement-sensitive parameters, whereas the QFI used by QEWC imposes a denser state-geometric constraint over the parameter space. 
By analyzing FI distributions, parameter displacements, and output distribution separability, we show that these distinct regularization geometries lead to distinct stability--plasticity behaviors during sequential training. 
This indicates that QEWC is not merely a direct quantum analogue of conventional EWC, but rather a state-geometric approach to understanding how quantum models preserve previously learned task representations.

In summary, our results demonstrate that QFI provides a physically motivated and geometrically meaningful metric for continual learning in quantum model. 
By characterizing parameter importance through the local sensitivity of the underlying quantum states, QEWC offers a complementary perspective to conventional CFI-based consolidation, which is tied to measurement-induced output statistics. 
Our results further indicate that this state-geometric viewpoint is particularly useful for analyzing forgetting under noisy quantum dynamics, where measurement-dependent FI can be strongly suppressed while the QFI retains a more stable sensitivity structure. 
At the same time, the practical effectiveness of QEWC depends on several implementation choices, including the estimation of the QFI, the use of diagonal approximations, the noise model, and the regularization strength controlling the stability--plasticity balance. 
These observations suggest that forgetting in quantum models should be studied not only through output-level performance, but also through the geometry of the parameterized quantum state manifold.

\section*{Acknowledgment}
Y.-C. Hsu and Y.-C. Lin thank the National Center
for High-Performance Computing (NCHC), National Institutes of Applied Research (NIAR), Taiwan, for providing computational and storage resources supported by the National Science and Technology Council (NSTC), Taiwan, under Grants No.
NSTC 114-2119-M-007-013. E.-J.Kuo acknowledges financial support from the National Science and Technology Council (NSTC) of Taiwan under Grant No.~NSTC~114-2112-M-A49-036-MY3.

\bibliographystyle{alpha}
\bibliography{ref}

\appendix
\renewcommand{\thesection}{Appendix~\Alph{section}.}

\section{Derivation of the EWC Objective}
\label{app:ewc-derivation}
\setcounter{equation}{0}
\renewcommand{\theequation}{A.\arabic{equation}}

In this appendix, we derive the elastic weight consolidation (EWC) objective from a Bayesian perspective. 
We consider a parametrized QNN with trainable parameters $\boldsymbol{\theta}$ that is trained sequentially on two tasks $T_1$ and $T_2$, associated with datasets $\mathcal{D}_1$ and $\mathcal{D}_2$, respectively. 
Within Bayesian continual learning, the goal is to infer the posterior distribution of the parameters after observing both datasets, namely $p(\boldsymbol{\theta} \mid \mathcal{D}_1,\mathcal{D}_2)$. 
Assuming that the datasets are conditionally independent given the model parameters, Bayes' rule gives
\begin{equation}
\log p(\boldsymbol{\theta}\mid\mathcal{D}_1,\mathcal{D}_2)
=
\log p(\mathcal{D}_2\mid\boldsymbol{\theta})
+
\log p(\boldsymbol{\theta}\mid\mathcal{D}_1)
-
\log p(\mathcal{D}_2\mid\mathcal{D}_1),
\label{eq:posterior_ewc1}
\end{equation}
where $\log p(\mathcal{D}_2 \mid \boldsymbol{\theta})$ is the log-likelihood of the data from task $T_2$, and $\log p(\boldsymbol{\theta} \mid \mathcal{D}_1)$ represents the posterior obtained after learning the first task. 
The last term, $\log p(\mathcal{D}_2\mid\mathcal{D}_1)$, is independent of $\boldsymbol{\theta}$ and therefore acts only as a normalization constant during optimization.

The negative log-likelihood of the new task is identified with the empirical loss, i.e.,
\[
-\log p(\mathcal{D}_2 \mid \boldsymbol{\theta})
=
\mathcal{L}_2(\boldsymbol{\theta}).
\]
Consequently, maximizing the posterior in Eq.~(\ref{eq:posterior_ewc1}) is equivalent to minimizing the loss on the new task while regularizing the parameters according to the posterior learned from the previous task.

The remaining term $\log p(\boldsymbol{\theta}\mid\mathcal{D}_1)$ encodes the information acquired from task $T_1$. 
However, the exact posterior distribution is generally intractable. 
Following the classical EWC framework, we approximate $p(\boldsymbol{\theta}\mid\mathcal{D}_1)$ by a Gaussian distribution centered at the optimal parameter vector $\boldsymbol{\theta}_1^{*}$ obtained after learning $T_1$. 
This corresponds to a second-order Taylor expansion of the log-posterior around $\boldsymbol{\theta}_1^{*}$:
\begin{equation}
\log p\left(\boldsymbol{\theta} \mid \mathcal{D}_1\right)
=
\log p\left(\boldsymbol{\theta}_1^{*} \mid \mathcal{D}_1\right)
+
\left.
\nabla_{\boldsymbol{\theta}}
\log p\left(\boldsymbol{\theta} \mid \mathcal{D}_1\right)
\right|_{\boldsymbol{\theta}=\boldsymbol{\theta}_1^{*}}^{\mathsf T}
(\boldsymbol{\theta}-\boldsymbol{\theta}_1^{*})
-
\frac{1}{2}
(\boldsymbol{\theta}-\boldsymbol{\theta}_1^{*})^{\mathsf T}
H_{\boldsymbol{\theta}_1^{*}}
(\boldsymbol{\theta}-\boldsymbol{\theta}_1^{*}),
\label{eq:taylor_ewc}
\end{equation}
where $H_{\boldsymbol{\theta}_1^{*}}$ denotes the negative Hessian matrix of $\log p(\boldsymbol{\theta}\mid\mathcal{D}_1)$ evaluated at $\boldsymbol{\theta}_1^{*}$. 
Since $\boldsymbol{\theta}_1^{*}$ maximizes the posterior after learning $T_1$, the first-order term vanishes. 
Substituting Eq.~(\ref{eq:taylor_ewc}) into Eq.~(\ref{eq:posterior_ewc1}) and discarding constants independent of $\boldsymbol{\theta}$, we obtain the regularized objective for learning the second task:
\begin{equation}
\mathcal{L}_{\mathrm{EWC}}(\boldsymbol{\theta})
=
\mathcal{L}_2(\boldsymbol{\theta})
+
\frac{1}{2}
(\boldsymbol{\theta}-\boldsymbol{\theta}_1^{*})^\top
H_{\boldsymbol{\theta}_1^{*}}
(\boldsymbol{\theta}-\boldsymbol{\theta}_1^{*}).
\label{eq:ewc_hessian}
\end{equation}

In practice, computing the full Hessian matrix is often computationally prohibitive. 
A common approximation is therefore to replace the Hessian with the CFI matrix, which provides a tractable measure of the local curvature around the optimum.

To make this approximation explicit while keeping track of the task dependence, we define the CFI matrix associated with task $T_1$ as $F^{(1)}_{\mathrm C}(\boldsymbol{\theta}_1^{*})$.
Assuming that the dataset $\mathcal{D}_1=\{(x,y)\}$ consists of independent samples drawn from the underlying data distribution, the likelihood factorizes as
\[
p(\mathcal{D}_1 \mid \boldsymbol{\theta})
=
\prod_{(x,y)\in\mathcal{D}_1}
p(y \mid x,\boldsymbol{\theta}).
\]
The matrix elements of $F^{(1)}_{\mathrm C}(\boldsymbol{\theta}_1^{*})$ are given by

\begin{equation}
\left[F^{(1)}_{\mathrm C}(\boldsymbol{\theta}_1^{*})\right]_{ij}
=
\mathbb{E}_{(x,y)\sim \mathcal{D}_1}
\left[
\frac{\partial \log p(y \mid x, \boldsymbol{\theta})}{\partial \theta_i}
\frac{\partial \log p(y \mid x, \boldsymbol{\theta})}{\partial \theta_j}
\right]_{\boldsymbol{\theta}=\boldsymbol{\theta}_1^{*}} .
\label{eq:fisher_def}
\end{equation}

Under standard regularity conditions, this task-dependent CFI matrix coincides with the negative expected Hessian of the log-likelihood,

\begin{equation}
F^{(1)}_{\mathrm C}(\boldsymbol{\theta}_1^{*})
=
-
\mathbb{E}
\left[
\nabla^2_{\boldsymbol{\theta}}
\log p(\mathcal{D}_1 \mid \boldsymbol{\theta})
\right]_{\boldsymbol{\theta}=\boldsymbol{\theta}_1^{*}} ,
\label{eq:fisher_hessian_relation}
\end{equation}

which justifies its use as a local curvature approximation around $\boldsymbol{\theta}_1^{*}$.
In the classical EWC approximation, the curvature contribution from the prior is either neglected or absorbed into the regularization strength, so that the posterior curvature is approximated by the Fisher information of the task likelihood.

Replacing the Hessian in Eq.~(\ref{eq:ewc_hessian}) with $F^{(1)}_{\mathrm C}(\boldsymbol{\theta}_1^{*})$ yields the EWC objective
\begin{equation}
\mathcal{L}_{\mathrm{EWC}}(\boldsymbol{\theta})
=
\mathcal{L}_2(\boldsymbol{\theta})
+
\frac{\lambda}{2}
(\boldsymbol{\theta}-\boldsymbol{\theta}_1^{*})^\top
F^{(1)}_{\mathrm C}(\boldsymbol{\theta}_1^{*})
(\boldsymbol{\theta}-\boldsymbol{\theta}_1^{*}),
\label{eq:ewc_fisher}
\end{equation}

where $\lambda$ controls the strength of the regularization. 
This penalty discourages updates along parameter directions that are important for the previously learned task.

For computational efficiency, only the diagonal elements of $F^{(1)}_{\mathrm C}(\boldsymbol{\theta}_1^{*})$ are often retained, giving

\begin{equation}
\mathcal{L}_{\mathrm{EWC}}(\boldsymbol{\theta})
=
\mathcal{L}_2(\boldsymbol{\theta})
+
\frac{\lambda}{2}
\sum_i
\left[F^{(1)}_{\mathrm C}(\boldsymbol{\theta}_1^{*})\right]_{ii}
(\theta_i-\theta_{1,i}^{*})^2.
\label{eq:ewc_diag}
\end{equation}

Here, $\left[F^{(1)}_{\mathrm C}(\boldsymbol{\theta}_1^{*})\right]_{ii}$ measures the importance of the $i$-th parameter for preserving the knowledge acquired from task $T_1$.

For a general previously learned task $T_j$, we denote the corresponding task-dependent CFI matrix by $F^{(j)}_{\mathrm C}(\boldsymbol{\theta}_j^{*})$, evaluated at the optimum of task $T_j$ and with respect to the data distribution $\mathcal{D}_j$.
Equivalently, its matrix elements are defined as

\begin{equation}
\left[F^{(j)}_{\mathrm C}(\boldsymbol{\theta}_j^{*})\right]_{mn}
=
\mathbb{E}_{(x,y)\sim \mathcal{D}_j}
\left[
\frac{\partial \log p(y \mid x, \boldsymbol{\theta})}{\partial \theta_m}
\frac{\partial \log p(y \mid x, \boldsymbol{\theta})}{\partial \theta_n}
\right]_{\boldsymbol{\theta}=\boldsymbol{\theta}_j^{*}} .
\label{eq:fisher_def_task_j}
\end{equation}

This notation makes explicit that each consolidation term uses the Fisher matrix associated with the corresponding previous task, rather than a task-independent curvature matrix.

The above construction naturally generalizes to a sequence of tasks $\{T_1,\ldots,T_k\}$. 
When learning task $T_k$, the EWC objective accumulates the Fisher-weighted penalties associated with all previously learned tasks:
\begin{equation}
\mathcal{L}_{\mathrm{EWC}}^{(k)}(\boldsymbol{\theta})
=
\mathcal{L}_k(\boldsymbol{\theta})
+
\frac{\lambda}{2}
\sum_{j=1}^{k-1}
(\boldsymbol{\theta}-\boldsymbol{\theta}_{j}^{*})^\top
F^{(j)}_{\mathrm C}(\boldsymbol{\theta}_{j}^{*})
(\boldsymbol{\theta}-\boldsymbol{\theta}_{j}^{*}),
\label{eq:ewc_general}
\end{equation}

where $\boldsymbol{\theta}_{j}^{*}$ denotes the optimal parameter vector obtained after training on task $T_j$, and $F^{(j)}_{\mathrm C}(\boldsymbol{\theta}_{j}^{*})$ denotes the CFI matrix evaluated at $\boldsymbol{\theta}_{j}^{*}$ using the data distribution of task $T_j$. 
Thus, EWC preserves previously acquired knowledge by penalizing changes along parameter directions that were important for earlier tasks.

When the task-dependent weighting factor $\alpha_j^{(k)}$ is introduced, the corresponding weighted EWC objective is obtained by multiplying each previous-task penalty by $\alpha_j^{(k)}$:
\begin{equation}
\mathcal{L}_{\mathrm{EWC,weighted}}^{(k)}(\boldsymbol{\theta})
=
\mathcal{L}_k(\boldsymbol{\theta})
+
\frac{\lambda}{2}
\sum_{j=1}^{k-1}
\alpha_j^{(k)}
(\boldsymbol{\theta}-\boldsymbol{\theta}_{j}^{*})^\top
F^{(j)}_{\mathrm C}(\boldsymbol{\theta}_{j}^{*})
(\boldsymbol{\theta}-\boldsymbol{\theta}_{j}^{*}) .
\label{eq:ewc_weighted_app}
\end{equation}
This weighted form is the CFI-based counterpart of the weighted QEWC objective used in the main text.

\section{Diagonal Approximation of the Quantum Fisher Matrix}

\begin{figure}[!t]
\centering
\includegraphics[width=0.82\textwidth]{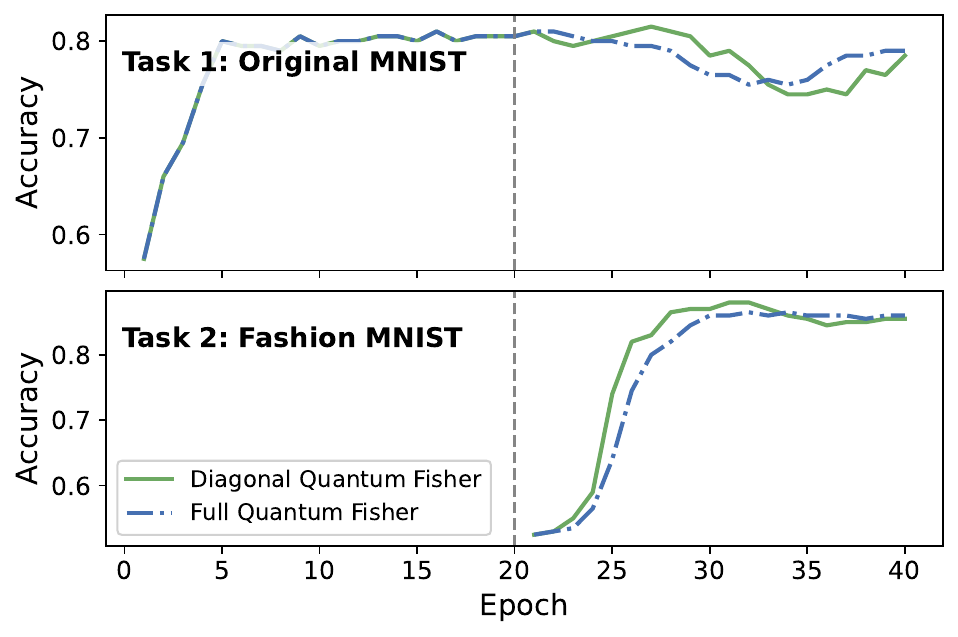}
\caption{\textbf{Comparison between diagonal-QFI and full-QFI QEWC.} Classification accuracy of the VQC trained sequentially on the original MNIST task and the Fashion-MNIST task using either the diagonal approximation to the QFI metric or the full QFI matrix in the QEWC regularization term. The dashed vertical line indicates the task boundary.}
\label{fig:full_diag}
\end{figure}

\setcounter{equation}{0}
\renewcommand{\theequation}{C.\arabic{equation}}
\label{app:full}

In our main numerical simulations, we employ a diagonal approximation of the QFI matrix within the QEWC framework. 
The full QFI matrix characterizes correlations among different parameter directions; however, its computational and memory requirements increase rapidly with the number of trainable parameters. 
This issue becomes particularly severe for deep variational quantum circuits, where the number of circuit parameters grows with both the number of qubits and the number of variational layers. 
Therefore, from a practical standpoint, we approximate the QFI matrix by retaining only its diagonal entries.

In the diagonal-QFI implementation, we construct the parameter-wise importance weights by evaluating the task-averaged QFI matrix and retaining only its diagonal entries for use in the QEWC penalty. 
The resulting diagonal regularizer preserves the local sensitivity associated with each individual parameter direction, while neglecting off-diagonal correlations between different parameters. 
This approximation reduces the computational and storage costs of representing and applying the full QFI matrix, while retaining the parameter-wise importance information required by the diagonal quadratic regularization term.

We further assess the effect of neglecting off-diagonal QFI elements by comparing the continual-learning performance of diagonal-QFI QEWC with that of full-QFI QEWC. 
Because the full QFI calculation incurs substantially higher computational cost, this comparison is performed using a reduced VQC with 10 variational layers. 
The model is trained sequentially on the original MNIST task and the Fashion-MNIST task under otherwise identical learning settings.

As shown in Fig.~\ref{fig:full_diag}, the diagonal-QFI and full-QFI variants exhibit similar learning trajectories across the two-task sequence. 
Both methods retain performance on the first task after the transition to Fashion-MNIST and achieve comparable accuracy on the second task. 
These results suggest that, for this benchmark, the diagonal approximation captures the dominant parameter-wise sensitivity required by the QEWC framework. 
Therefore, we adopt the diagonal QFI matrix in the main numerical simulations.

\section{Proof of the Local QFI Stability Bounds}
\setcounter{equation}{0}
\renewcommand{\theequation}{B.\arabic{equation}}
\label{app:qewc-bound}

In this appendix, we provide the proof of the local bounds stated in Sec.~\ref{sec:qfi_theoretical_justification}. 
The proof proceeds in two steps. 
First, we use the local Bures-metric interpretation of the QFI to show that the QFI-weighted displacement controls the drift of the quantum states and the induced output probabilities. 
Second, under a mild nonzero-probability condition, we show that this probability-drift control implies a local bound on the increase of the previous-task cross-entropy loss.

\paragraph{Proof of the state- and probability-drift bound.}

Let $\Delta\boldsymbol{\theta}=\boldsymbol{\theta}-\boldsymbol{\theta}_j^\ast$ denote a sufficiently small displacement from the previous-task optimum. 
By the definition of the QFI as the local metric associated with the Bures distance, for each input $\mathbf{x}$ we have
\begin{equation}
D_{\mathrm{Bures}}^2
\left(
\rho_{\mathbf{x}}(\boldsymbol{\theta}_j^\ast),
\rho_{\mathbf{x}}(\boldsymbol{\theta}_j^\ast+\Delta\boldsymbol{\theta})
\right)
=
\frac{1}{4}
\Delta\boldsymbol{\theta}^{\mathsf T}
F_Q
\left[
\rho_{\mathbf{x}}(\boldsymbol{\theta}_j^\ast)
\right]
\Delta\boldsymbol{\theta}
+
o\!\left(\|\Delta\boldsymbol{\theta}\|^2\right),
\label{eq:single_input_bures_expansion_app}
\end{equation}
where we use the convention that the QFI is four times the local Bures metric. 
Averaging Eq.~\eqref{eq:single_input_bures_expansion_app} over $(\mathbf{x},y)\in\mathcal{D}_j$ gives
\begin{equation}
\frac{1}{|\mathcal{D}_j|}
\sum_{(\mathbf{x},y)\in\mathcal{D}_j}
D_{\mathrm{Bures}}^2
\left(
\rho_{\mathbf{x}}(\boldsymbol{\theta}_j^\ast),
\rho_{\mathbf{x}}(\boldsymbol{\theta}_j^\ast+\Delta\boldsymbol{\theta})
\right)
=
\frac{1}{4}
\Delta\boldsymbol{\theta}^{\mathsf T}
\bar{F}_{Q}^{(j)}(\boldsymbol{\theta}_j^\ast)
\Delta\boldsymbol{\theta}
+
o\!\left(\|\Delta\boldsymbol{\theta}\|^2\right),
\label{eq:qfi_bures_local_expansion_app}
\end{equation}
which proves the local Bures-distance relation.

We next relate this state-level stability to output-probability stability.
The trace distance between two quantum states is controlled by their
Bures distance. Therefore, for sufficiently small $\Delta\boldsymbol{\theta}$,
\begin{equation}
D_{\mathrm{tr}}
\left(
\rho_{\mathbf{x}}(\boldsymbol{\theta}_j^\ast),
\rho_{\mathbf{x}}(\boldsymbol{\theta}_j^\ast+\Delta\boldsymbol{\theta})
\right)
=
O
\left(
D_{\mathrm{Bures}}
\left(
\rho_{\mathbf{x}}(\boldsymbol{\theta}_j^\ast),
\rho_{\mathbf{x}}(\boldsymbol{\theta}_j^\ast+\Delta\boldsymbol{\theta})
\right)
\right).
\label{eq:bures_trace_relation_app}
\end{equation}
For each fixed readout observable $\hat Z_c$, the corresponding score
drift satisfies
\begin{align}
&
\left|
z_c(\mathbf{x},\boldsymbol{\theta}_j^\ast+\Delta\boldsymbol{\theta})
-
z_c(\mathbf{x},\boldsymbol{\theta}_j^\ast)
\right|
\nonumber\\
&\quad =
\left|
\mathrm{Tr}\!\left[
\hat Z_c
\left(
\rho_{\mathbf{x}}(\boldsymbol{\theta}_j^\ast+\Delta\boldsymbol{\theta})
-
\rho_{\mathbf{x}}(\boldsymbol{\theta}_j^\ast)
\right)
\right]
\right|
\nonumber\\
&\quad \le
\|\hat Z_c\|_{\infty}
\left\|
\rho_{\mathbf{x}}(\boldsymbol{\theta}_j^\ast+\Delta\boldsymbol{\theta})
-
\rho_{\mathbf{x}}(\boldsymbol{\theta}_j^\ast)
\right\|_1 .
\label{eq:score_trace_bound_app}
\end{align}
Since $\|\hat Z_c\|_{\infty}=1$ for Pauli-$Z$ observables, the score
drift is controlled by the trace distance. Moreover, because the
softmax map is Lipschitz continuous on finite-dimensional score vectors,
there exists a constant $L_{\mathrm{sm}}>0$ such that
\begin{equation}
\sum_c
\left|
p_{\boldsymbol{\theta}'}(c|\mathbf{x})
-
p_{\boldsymbol{\theta}}(c|\mathbf{x})
\right|
\le
L_{\mathrm{sm}}
\sum_c
\left|
z_c(\mathbf{x},\boldsymbol{\theta}')
-
z_c(\mathbf{x},\boldsymbol{\theta})
\right| .
\label{eq:softmax_lipschitz_app}
\end{equation}
Combining Eqs.~\eqref{eq:single_input_bures_expansion_app},
\eqref{eq:bures_trace_relation_app}, \eqref{eq:score_trace_bound_app},
and \eqref{eq:softmax_lipschitz_app}, we obtain, for each input
$\mathbf{x}$,
\begin{equation}
\sum_c
\left|
p_{\boldsymbol{\theta}_j^\ast+\Delta\boldsymbol{\theta}}(c|\mathbf{x})
-
p_{\boldsymbol{\theta}_j^\ast}(c|\mathbf{x})
\right|
=
O\!\left(
\sqrt{
\Delta\boldsymbol{\theta}^{\mathsf T}
F_Q\!\left[
\rho_{\mathbf{x}}(\boldsymbol{\theta}_j^\ast)
\right]
\Delta\boldsymbol{\theta}
}
\right)
+
o\!\left(\|\Delta\boldsymbol{\theta}\|\right).
\label{eq:single_input_probability_drift_app}
\end{equation}
Averaging Eq.~\eqref{eq:single_input_probability_drift_app} over
$(\mathbf{x},y)\in\mathcal{D}_j$ and using Jensen's inequality for the
concave square-root function gives
\begin{align}
&
\frac{1}{|\mathcal{D}_j|}
\sum_{(\mathbf{x},y)\in\mathcal{D}_j}
\sum_c
\left|
p_{\boldsymbol{\theta}_j^\ast+\Delta\boldsymbol{\theta}}(c|\mathbf{x})
-
p_{\boldsymbol{\theta}_j^\ast}(c|\mathbf{x})
\right|
\nonumber\\
&\quad =
O\!\left(
\frac{1}{|\mathcal{D}_j|}
\sum_{(\mathbf{x},y)\in\mathcal{D}_j}
\sqrt{
\Delta\boldsymbol{\theta}^{\mathsf T}
F_Q\!\left[
\rho_{\mathbf{x}}(\boldsymbol{\theta}_j^\ast)
\right]
\Delta\boldsymbol{\theta}
}
\right)
+
o\!\left(\|\Delta\boldsymbol{\theta}\|\right)
\nonumber\\
&\quad \le
O\!\left(
\sqrt{
\frac{1}{|\mathcal{D}_j|}
\sum_{(\mathbf{x},y)\in\mathcal{D}_j}
\Delta\boldsymbol{\theta}^{\mathsf T}
F_Q\!\left[
\rho_{\mathbf{x}}(\boldsymbol{\theta}_j^\ast)
\right]
\Delta\boldsymbol{\theta}
}
\right)
+
o\!\left(\|\Delta\boldsymbol{\theta}\|\right)
\nonumber\\
&\quad =
O\!\left(
\sqrt{
\Delta\boldsymbol{\theta}^{\mathsf T}
\bar{F}_{Q}^{(j)}(\boldsymbol{\theta}_j^\ast)
\Delta\boldsymbol{\theta}
}
\right)
+
o\!\left(\|\Delta\boldsymbol{\theta}\|\right).
\label{eq:qfi_probability_drift_bound_app}
\end{align}
This proves the probability-drift bound.
\hfill$\square$

\paragraph{Proof of the cross-entropy bound.}
Assume that the correct-label probabilities at the previous-task optimum are bounded away from zero, i.e., there exists $p_{\min}>0$ such that
\begin{equation}
p_{\boldsymbol{\theta}_j^\ast}(y|\mathbf{x})
\ge
p_{\min},
\qquad
\forall\,(\mathbf{x},y)\in\mathcal{D}_j .
\label{eq:pmin_assumption_app_1}
\end{equation}
The binary cross-entropy loss for the previous task is
\begin{equation}
\mathcal{L}_j(\boldsymbol{\theta})
=
-
\frac{1}{|\mathcal{D}_j|}
\sum_{(\mathbf{x},y)\in\mathcal{D}_j}
\log
p_{\boldsymbol{\theta}}(y|\mathbf{x}) .
\label{eq:previous_task_bce_app}
\end{equation}
Under the condition in Eq.~\eqref{eq:pmin_assumption_app_1}, and for sufficiently small $\Delta\boldsymbol{\theta}$, the correct-label probabilities remain bounded away from zero. 
Therefore, $-\log p$ is locally Lipschitz on the relevant interval. 
In particular, for sufficiently small $\Delta\boldsymbol{\theta}$, we have
$p_{\boldsymbol{\theta}_j^\ast+\Delta\boldsymbol{\theta}}(y|\mathbf{x})
\ge p_{\min}/2$, so $-\log p$ is Lipschitz on $[p_{\min}/2,1]$ with Lipschitz constant at most $2/p_{\min}$. 
Hence, there exists a task-dependent constant $C_j>0$ such that
\begin{align}
&
\mathcal{L}_j
\left(
\boldsymbol{\theta}_j^\ast+\Delta\boldsymbol{\theta}
\right)
-
\mathcal{L}_j
\left(
\boldsymbol{\theta}_j^\ast
\right)
\nonumber\\
&\quad \le
\frac{C_j}{|\mathcal{D}_j|}
\sum_{(\mathbf{x},y)\in\mathcal{D}_j}
\left|
p_{\boldsymbol{\theta}_j^\ast+\Delta\boldsymbol{\theta}}(y|\mathbf{x})
-
p_{\boldsymbol{\theta}_j^\ast}(y|\mathbf{x})
\right|
\nonumber\\
&\quad \le
O
\left(
\sqrt{
\Delta\boldsymbol{\theta}^{\mathsf T}
\bar{F}_{Q}^{(j)}(\boldsymbol{\theta}_j^\ast)
\Delta\boldsymbol{\theta}
}
\right)
+
o\!\left(\|\Delta\boldsymbol{\theta}\|\right),
\label{eq:qfi_bce_local_bound_app}
\end{align}
where the final step follows from Eq.~\eqref{eq:qfi_probability_drift_bound_app}. 
This proves the local control of the previous-task cross-entropy loss. 
\hfill$\square$

\section{Quantum Data Generation}
\setcounter{equation}{0}
\renewcommand{\theequation}{C.\arabic{equation}}
\label{app:data}

To evaluate the ability of the quantum classifier to recognize quantum many-body states in $T_3$, we consider a binary classification task involving ground states sampled from parameter regimes associated with the symmetry-protected topological (SPT) and antiferromagnetic (ATF) phases~\cite{Zhang2026Experimental, Phase2011}. The quantum states are generated from a one-dimensional cluster-Ising spin chain under open boundary conditions, governed by the Hamiltonian:
% \en{phase transition only occurs at infinite $N$, we can say
% For $N=4$, these should be understood as finite-size ground states sampled from parameter regimes corresponding to the SPT and ATF phases in the thermodynamic limit.}

\begin{equation}
H(h)
=
-\sum_{j=2}^{N-1}
\hat{\sigma}^{x}_{j-1}
\hat{\sigma}^{z}_{j}
\hat{\sigma}^{x}_{j+1}
+
h\sum_{j=1}^{N-1}
\hat{\sigma}^{y}_{j}
\hat{\sigma}^{y}_{j+1},
\label{eq:cluster_ising_hamiltonian}
\end{equation}

where $N=4$ is the number of qubits, $h$ controls the relative strength of the nearest-neighbor interaction, and
$\hat{\sigma}^{l}_{j}$, with $l\in\{x,y,z\}$, denotes the corresponding Pauli operator acting on the $j$-th qubit.

In the thermodynamic limit, the cluster-Ising model exhibits a quantum phase transition at $h=1$, separating the SPT cluster phase for $h<1$ from the ATF phase for $h>1$. Since our numerical simulations employ a finite system with $N=4$, no sharp phase transition occurs. Accordingly, the generated states should be interpreted as finite-size ground states sampled from parameter regimes corresponding to the two phases in the thermodynamic limit.

To construct the quantum dataset, we sample the interaction parameter $h$ from two well-separated intervals away from the critical region. Specifically, samples with
$h\in[0.0,0.5]$ are assigned to the SPT class, whereas samples with
$h\in[2.5,3.0]$ are assigned to the ATF class~\cite{Zhang2026Experimental}. For each sampled value of $h$, we construct the sparse matrix representation of $H(h)$ and obtain its lowest-energy eigenvector through exact diagonalization. The resulting normalized ground-state vector is directly supplied to the quantum classifier as the input quantum state.

\end{document}